# MedTQ: Dynamic Topic Discovery and Query Generation for Medical Ontologies

**Feichen Shen, PhD, Yugyung Lee, PhD**
**School of Computing and Engineering**
**University of Missouri-Kansas City, MO, USA**

***Abstract***

Background

Biomedical ontology refers to a shared conceptualization for a biomedical domain of interest that has vastly improved data management and data sharing through the open data movement. The rapid growth and availability of biomedical data make it impractical and computationally expensive to perform manual analysis and query processing with the large scale ontologies. The lack of ability in analyzing ontologies from such a variety of sources, and supporting knowledge discovery for clinical practice and biomedical research should be overcome with new technologies.

Methods

In this study, we developed a Medical Topic discovery and Query generation framework (MedTQ), which was composed by a series of approaches and algorithms. A predicate neighborhood pattern-based approach introduced has the ability to compute the similarity of predicates (relations) in ontologies. Given a predicate similarity metric, machine learning algorithms have been developed for automatic topic discovery and query generation. The topic discovery algorithm, called the hierarchical K-Means algorithm was designed by extending an existing supervised algorithm (K-means clustering) for the construction of a topic hierarchy. In the hierarchical K-Means algorithm, a level-by-level optimization strategy was selected for consistent with the strongly association between elements within a topic. Automatic query generation was facilitated for discovered topic that could be guided users for interactive query design and processing.

Results

A topic hierarchy was constructed using the DrugBank ontology as a case study. In this hierarchy, 8 specific topics were generated. Ranking of predicates and concepts of these topics were also computed. An experiment has been conducted to find an optimal number of the topics using the four different clustering algorithms, K-means, Clustering Large Application (Clara), Partition Around Medoids (Pam), and Hierarchical Clustering. A number of SPARQL queries generated were automatically generated from the discovered topics to demonstrate the ability to retrieve information from the DrugBank ontology.

Conclusions

The paper addresses knowledge discovery through analysis of ontologies for clinical practice and biomedical research. The model of predicate-oriented neighborhood pattern is explained in the context of topic discovery and query generation for ontologies. The MedTQ framework enhances knowledge discovery by capturing underlying structures from domain specific data and ontologies.

## I. INTRODUCTION

In recent years, a large number of ontologies have been introduced for clinical practice and biomedical research. As a result, there have been increasing demands on knowledge discovery and sharing of large scale biomedical data. The first notable effort shown by the biomedical and scientific community toward connecting scattered medical data is to materialize them through the open data movement (i.e., the Linked Data, bio2RDF, OBO, LinkedCT)[1]. However, many of these ontologies are still not fully annotated nor connected with other ontologies. The rapid growth and availability of biomedical data make it impractical and computationally expensive to perform manual analysis and query processing with the large scale ontologies[2].

In reality, there is a limited capacity to carry out dynamic analysis and query processing with these large-scale datasets. The current medical ontologies and services are not sufficient to be combined together due to the lack of underlying cohesive structure and semantics[3]. The previous work on inferring ontology structure[4] have mainly focused on determining whether or not there was a relationship between a given pair of concepts irrespective of the connections between them or the strength of the association. In particular, the techniques have yet to be fully implemented on dynamic analysis and query processing with the ontologies. There are endpoint services (e.g., BioPortal) for biomedical research[5], but most of them are not functioning properly or even if they are working, specific query content and formats are not at a practical level.

Knowledge discovery through analysis of ontologies for clinical practice and biomedical research has become a challenging task[6, 7]. We need an advanced approach to thoroughly understand ontologies instead of simply getting a slice of reference ontology and applying them for a query process or decision support[8]. Subsequently, it is essential to know what information exists and what meaningful relationships are present among associated domains (e.g., identification of genes responsible for a disease[9, 10], development of drugs for their treatment[11] or detect associations between diseases and phenotypes[12, 13]). Once the structure of an ontology has been defined, it is useful to identify and differentiate the context and strength of influence in domains and extract cohesive structure and semantics from ontologies.

In this paper, we presented a semantic framework, called the *MedTQ* framework. The *MedTQ* performs dynamic topic discovery (relationships) and automatic query generation through the analysis of predicates among concepts and role names, called the Predicate Neighborhood Patterns (PNP) in biomedical ontologies. Furthermore, a new clustering technique, called the Hierarchical Predicate-based K-Means clustering (HPKM) was proposed to dynamically identify latent topics and automatically generate queries based on the discovered patterns. We have also implemented an interactive tool that allows researchers to explore ontologies and generate queries by combining interesting contents, and then retrieve relevant information in a logical way. In addition, topics were further evaluated based on prioritized information of medical ontologies for biomedical research.

The contributions of this paper are fourfold.

- Formal definition of predicate neighborhood patterns (PNP)
- Hierarchical Predicate-based K-means clustering (HPKM)
- Automatic query generation based on the discovered topics (clusters)
- Interactive tool for dynamic query generation and an endpoint for query processing

A case study was designed with a major medical ontology (i.e., DrugBank[14]) to demonstrate the dynamic topic discovery and query generation by the *MedTQ* framework. We have implemented a prototype of the *MedTQ* system and evaluated the statistical significance of our model in discovery of topics. In addition, we successfully validated the clustering results, thereby providing a solid evidence for automatic query generation.

The major content of this paper is organized as follows: We first present the *MedTQ* framework in Section II. We then describe the implementation of the *MedTQ* system in Section III. We present the main results and discussion in Section IV and Section V. The conclusion is discussed in Section VI.

## II. METHODS

In this paper, we proposed a semantic framework, called the *MedTQ* that identifies the relationships present among concepts and discovers knowledge through the construction of a hierarchy of topics (called the *topic hierarchy*) from biomedical ontologies. In the *topic hierarchy*, the abstractions of topics are analyzed for preserving information that is relevant in a given context (topic) without revealing the details of an underlying ontology structure. The topic models based on the relationships and their neighborhood patterns are defined as a graph in different levels of abstraction.

We first rationalized a predicate-centric model '*Predicate Neighboring Patterns* (PNP)' that specifies high connectivity on the RDF/OWL graph for information sharing. Second, we presented a Hierarchical Predicate-based K-Means clustering (HPKM) algorithm to cluster the graph based on the PNP patterns. Finally, we presented a query generation model for automatic query generation from the discovered topics.

A. *Predicate Neighboring Patterns*

The predicate neighboring patterns (PNP) defines the patterns of predicates playing an important role in sharing information and connecting the concepts in the ontology. The RDF/OWL data model[15] specifies resources (information on the entities and their relationship in the given ontologies) in the form of triples <subject (S), predicate (P), object (O)>, where S denotes the resource, and P denotes aspects of the resource and expresses a relationship between S and O. Multiple S can be connected to multiple O through a single predicate. A predicate P is representing a binary relation between two concepts (S and O) in ontologies. In RDF/OWL, P is represented as a property to express a kind of relationship (e.g., rdfs:subClassOf) between domain (subject) and range (object). The subject and object can be either from the same ontology or from different ontologies. From the basic unit of <S, P, O>, a specific context of a predicate P can be discovered from the associated concepts (S and O). Interestingly, the neighbors of predicates P will also provide additional information through the association context.

In this paper, two types of predicate patterns are defined as follows:

***Share Pattern***: As shown in Figure 1, this pattern describes the resources sharing relationships (P) between interacting concepts such as shared subjects (S) or shared objects (O) through the given relationship. Assume that two predicates are given as follows: $P_1 <S_i, O_i>$ and $P_2 <S_j, O_j>$ where $S_i$, $S_j$ are a set of subjects and $O_i$, $O_j$ are a set of objects in given ontologies. The pattern describes that the same subject and object are shared by two predicates $P_1$ and $P_2$, the same subject shared, and the object shared.

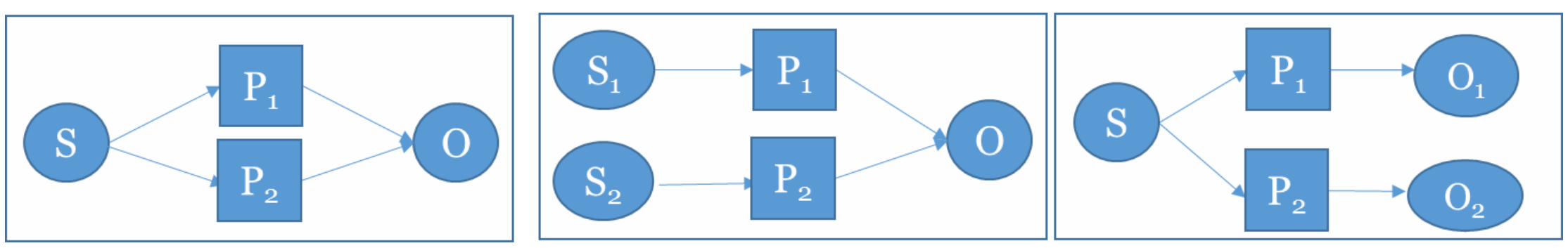

**Figure 1** Predicate Sharing Patterns

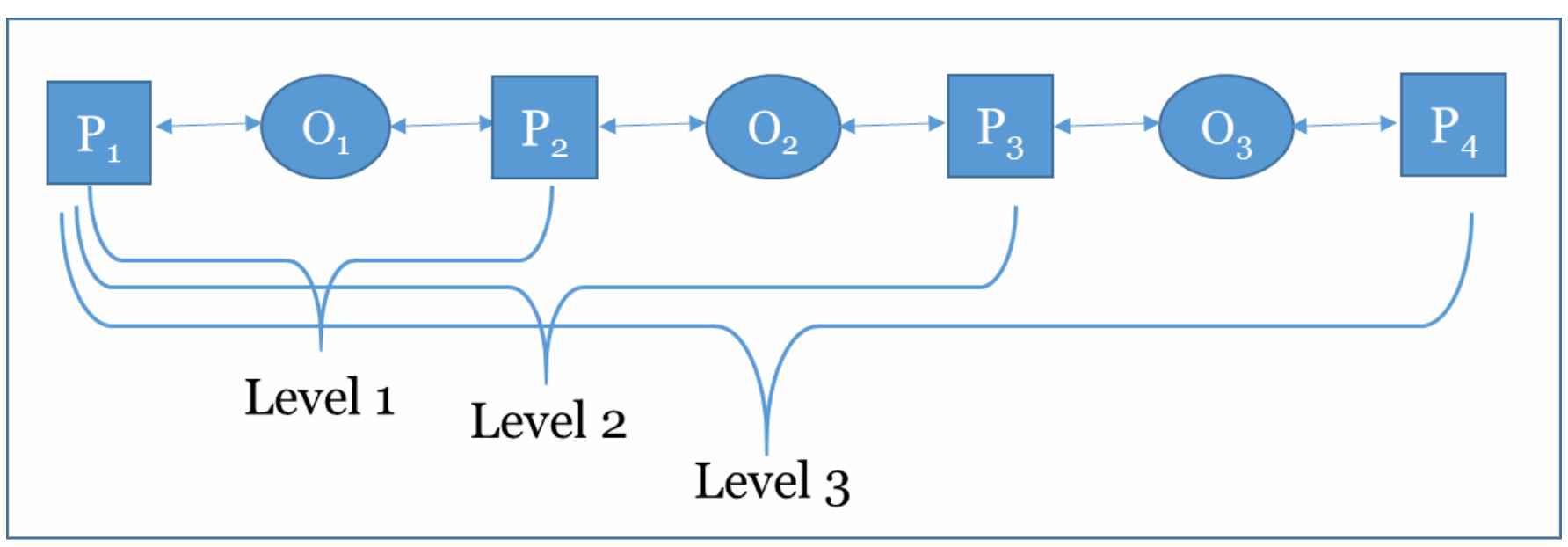

**Figure 2** Predicate Connection Patterns

***Connection Pattern***: As shown in Figure 2, this pattern describes the relationships based mainly on the connectivity of concept(s) through the respective predicates. This pattern is a frequently recurring pattern with predicates observed during query processing as the basis for joining one query pattern to another. This type of pattern describes the comprehension of the connectivity relationships between interacting predicates. Assume that two predicates are given as follows: $P_1 <S_i, O_i>$ and $P_2 <S_j, O_j>$ where $O_i$ is equal *to* $S_j$ and $P_1$ is directly connected to $P_2$ in the given ontologies $O_i$, $O_j$. Since the connection pattern at level 1 will be modelled in Shared Patterns, Connection Patterns are restricted to any patterns whose levels are greater than or equal to 2.

B. *Predicate Neighboring Measurements*

We define the measurement for the predicate neighboring patterns (PNP) in terms of sets of concepts and relations (predicates) over the ontologies. For this purpose, we now describe how to quantify similarities between different predicates based on the PNP pattern describing the relationships between predicates $\boldsymbol{P_i}$ and $\boldsymbol{P_j}$ through a concept $\boldsymbol{C}$. We formally define the similarity between predicates based on the shared patterns and connection patterns.

**Definition 1:** Given a directed graph $G(C, P)$, concepts $C$ denote subject S and object O and $P$ predicate in a RDF schema graph, respectively. Let $d(P_i, P_j)$ represent the number of concepts $C$ between $P_i$ and $P_j$. $r(P_i, P_j)$ determines if a predicate $P_i$ is reachable from another predicate $P_j$. $l(P_i, P_j)$ indicates the shortest distance between $P_i$ and $P_j$.

$$l(P_i, P_j) = \begin{cases} 0, & P_i = P_j \\ 1, & d(P_i, P_j) = 1 \\ L_1 + L_2, & L_1 = d(P_i, P_k)\ L_2 = d(P_k, P_j) \\ & r(P_i, P_k) = true, r(P_k, P_j) = true, \\ & r(P_i, P_j) = true \end{cases}$$

The similarity measurement for the PNP patterns varies based on different neighboring levels for each pair of predicates. Basically, we gave a higher shared score to predicates with more shared concepts and lower scores to predicates with less shared ones. Similarly, we gave a higher connection similarity score to closer predicates and lower scores to further predicates. We now define these two probability based similarity scores: i) $\boldsymbol{PS_s(P_i, P_j)}$ is defined as a shared pattern of any two predicates $\boldsymbol{P_i}$ and $\boldsymbol{P_j}$ ii) $\boldsymbol{PS_c(P_i, P_j)}$ for a connection pattern of any two predicates.

**Definition 2:** Given predicates $P_i$ and $P_j$ in a directed RDF schema Graph $G(C, P)$. Let $C(P_i)$ and $C(P_j)$ denote the entities (subjects or objects) that are directed connected to $P_i$ and $P_j$ regardless of the direction respectively. $PS_s(P_i, P_j)$ indicates the probability-based similarity for a *shared pattern* between $P_i$ and $P_j$.

$$PS_s(P_i, P_j) = \begin{cases} 1, & l(P_i, P_j) = 0 \\ 0, & l(P_i, P_j) \to \infty \\ \dfrac{(|C(P_i) \cap C(P_j)|)^2}{|C(P_i)| * |C(P_j)|}, & Otherwise \end{cases}$$

**Definition 3:** For a connection pattern of any two predicates $P_i$ and $P_j$, $PS_c(P_i, P_j)$ defines the probability-based similarity for a connection pattern between $P_i$ and $P_j$ as follows:

$$PS_c(P_i, P_j) = \begin{cases} PS_s(P_i, P_k) * PS_s(P_k, P_j), & l(P_i, P_j) = 2 \\ \max\limits_{i \le k < j}\{ PS_c(P_i, P_k) * PS_c(P_k, P_j)\}, & l(P_i, P_j) > 2 \end{cases}$$

The definition is influenced by the chain matrix multiplication problem (a kind of dynamic programming) that involves the question of determining the optimal sequence for performing a series of operations. After we got the similarity score for all pairs of predicates, we used formula in Definition 4 and 5 to generate a similarity matrix for clustering.

**Definition 4:** Given the total number of predicate *n* and the probability-based similarity score for shared patterns $PS_c(P_i, P_j)$ and connection patterns $PS_s(P_i, P_j)$ between predicates $P_i$ and $P_j$, $\mathrm{SM}[P_i, P_j]$ indicates a similarity matrix for all pairs of predicates $P_i$ and $P_j$

$$\mathrm{SM}[P_i, P_j] = \begin{cases} PS_c(P_i, P_j), & l(P_i, P_j) \ge 2 \\ PS_s(P_i, P_j), & Otherwise \end{cases}$$

As shown in Figure 3, an example of the predicate similarity computation for shared patterns and connection patterns was presented. In this example, a shared pattern is identified between predicates $P_1$ and $P_2$ and connection patterns are identified between $P_1$ and $P_3$, $P_1$ and $P_4$, $P_1$ and $P_5$. Based on the PNP patterns, $\mathrm{SM}[P_i, P_j]$ is computed.

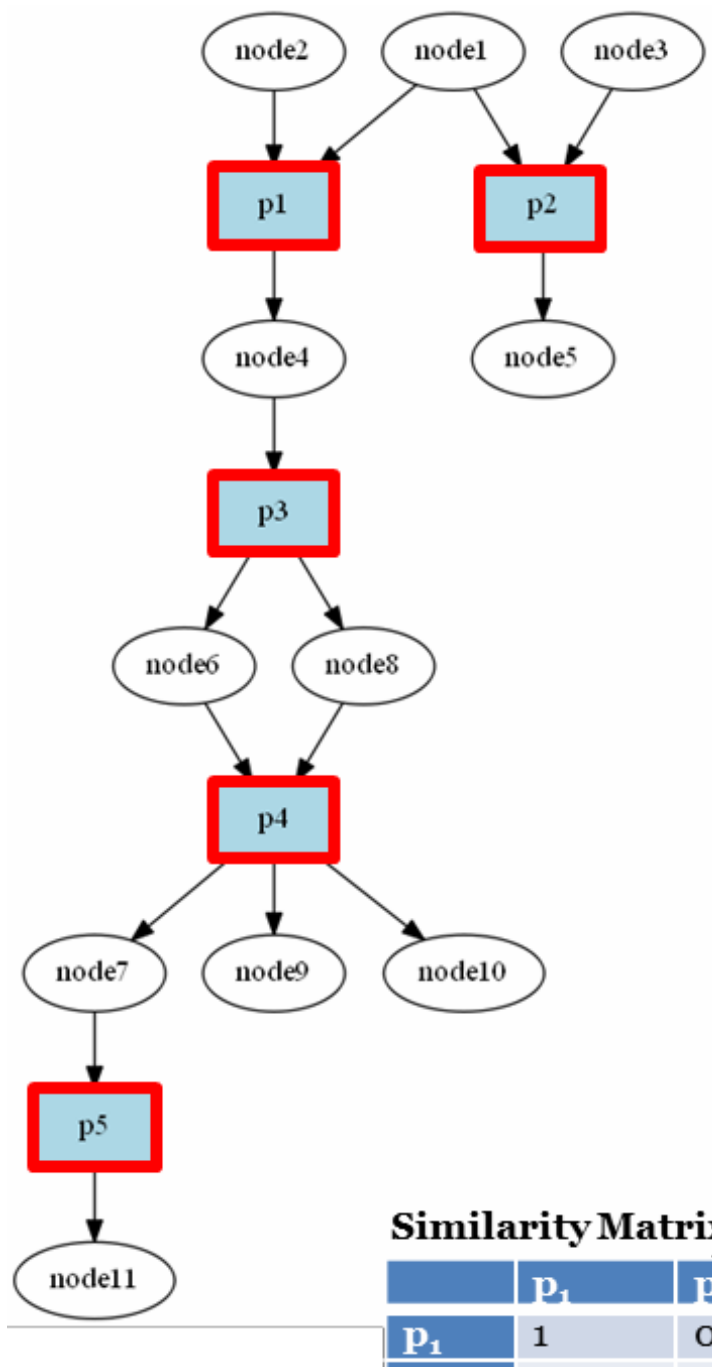

**Level1: ($P_1$, $P_2$), ($P_1$, $P_3$)**

$PS_s(P_1, P_2) = \frac{1}{3*3} = \frac{1}{9}$ =0.11

$PS_s(P_1, P_3) = \frac{1}{3*3} = \frac{1}{9}$ =0.11

$PS_s(P_3, P_4) = \frac{2^2}{3*5} = \frac{4}{15}$ =0.27

$PS_s(P_4, P_5) = \frac{1}{5*2} = \frac{1}{10}$ =0.1

**Level2: ($P_1$, $P_4$)**

$PS_c(P_1, P_4)$
$= PSs(P_1, P_3) * PSs(P_3, P_4)$
$= \frac{1}{9} * \frac{4}{15} = \frac{4}{135}$ =0.03

$PS_c(P_3, P_5) = PS_s(P_3, P_4) * PSs(P_4, P_5)$
$= \frac{4}{15} * \frac{1}{10} = \frac{4}{150}$ = 0.026

**Level3: ($P_1$, $P_5$)**

$PS_c(P_1, P_5) =$
$Max(PS_c(P_1, P_4) * PSs(P_4, P_5),$
$PS_s(P_1, P_3) * PSc(P_3, P_5)) =$
$Max(\frac{4}{135} * \frac{1}{10}, \frac{1}{9} * \frac{4}{150})$=0.0029

$PS_c(P_2, P_3) = 0$
$PS_c(P_2, P_4)$= 0
$PS_c(P_2, P_5)) = 0$

**Similarity Matrix (SM)**

| | $P_1$ | $P_2$ | $P_3$ | $P_4$ | $P_5$ |
|---|---|---|---|---|---|
| $P_1$ | 1 | 0.11 | 0.11 | 0.03 | 0.0029 |
| $P_2$ | 0.11 | 1 | 0 | 0 | 0 |
| $P_3$ | 0.11 | 0 | 1 | 0.27 | 0.026 |
| $P_4$ | 0.03 | 0 | 0.27 | 1 | 0.2 |
| $P_5$ | 0.0029 | 0 | 0.026 | 0.2 | 1 |

**Figure 3** Predicate Neighboring Patterns (PNP) and Similarity Matrix

### C. *Hierarcical Predicate-based K-Means Clustering*

The clustering approach we proposed here is based on the similarity measurement for the predicate neighboring patterns (PNP) inherent in the ontologies. We posited that predicate-based clustering is a required step for efficient query processing involving the alignment and integration of ontologies. Given that predicates are more closely related to some predicates than others, predicates can be clustered for efficient query processing - the task of classifying a collection of predicates into clusters (or topics). The guiding principle is to minimize inter-cluster (inter-topic) similarity and maximize intra-cluster (intra-topic) similarity, based on the similarity measure for the PNP patterns.

We now present our clustering algorithm, called the Hierarchical Predicate-based K-Means clustering (HPKM) that is designed by combining the divisive hierarchical clustering algorithm[16] and K-Means algorithms[17] for generating *K* topics level-by-level in an optimal manner. Similar to the K-Means algorithm, the HPKM is an unsupervised learning approach partitioning ontologies into *K* topics by clustering each predicate in the ontologies with the nearest mean. Similar to the divisive hierarchical clustering algorithm[16], the HPKM clusters ontologies into smaller topics in a hierarchical manner. The PNP pattern-based similarity and the silhouette width (SW)[18] were computed for achieving the objective of the clustering which is maximizing intra-cluster similarities and minimizing inter-cluster similarities[19]. If the SW of a topic is higher than α, this topic will be clustered into K smaller topics. The value of silhouette $sw(p_i)$ (i.e., silhouette width) can be ranged between -1 and 1. For each predicate $p_i$, we computed the following two similarity: inter-cluster similarity and intra-cluster similarity.

Intra-cluster similarity $\boldsymbol{a}(\boldsymbol{p_i})$: This measure refers to the similarity of data in a single cluster. Let $\boldsymbol{a}(\boldsymbol{p_i})$ be the average dissimilarity of $\boldsymbol{p_i}$ (taking the inverse of the SM matrix computed from the PNP algorithm) with all other data within the same cluster. It can be validated how well $\boldsymbol{p_i}$ is assigned to its cluster according to $\boldsymbol{a}(\boldsymbol{p_i})$ such as the smaller the value, the better the assignment. We then define the average dissimilarity of predicate $\boldsymbol{p_i}$ to a cluster $\boldsymbol{C}$ as the average of the distance from $\boldsymbol{p_i}$ to predicates in $\boldsymbol{C_i}$.

Inter-cluster similarity $\boldsymbol{b}(\boldsymbol{p_i})$: This measure refers to the similarity between clusters. Let $\boldsymbol{b}(\boldsymbol{p_i})$ be the lowest average similarity of $\boldsymbol{p_i}$ to the sibling clusters $\boldsymbol{C_j}$ that has the same parent cluster with $\boldsymbol{C_i}$ of which $\boldsymbol{p_i}$ is not a member. The cluster with this lowest average similarity is said to be the "sibling (neighboring) cluster", $\boldsymbol{C_j}$, of $\boldsymbol{p_i}$ because it is the next best fit cluster for predicate $\boldsymbol{p_i}$.

A silhouette width can be computed as follows:

$$sw(p_i) = \frac{b(p_i) - a(p_i)}{\max\{a(p_i), b(p_i)\}}$$

More specifically, it can be defined as follows: There are three possible cases about the silhouette width: (i) if the silhouette width $sw(p_i)$ is close to one, this means that the predicate $p_i$ is appropriately clustered. (ii) If $sw(p_i)$ is close to a negative one, then the predicate p would be not appropriate here but would be more appropriate if it was clustered in its neighboring cluster $C_j$. (iii) If $sw(p_i)$ is near zero then this means that the predicate $p_i$ is on the border of two natural clusters, namely $C_i\ and\ C_j$.

$$sw(p_i) = \begin{cases} 1 - \dfrac{a(p_i)}{b(p_i)}, & if\ a(p_i) < b(p_i) \\ 0, & if\ a(p_i) = b(p_i) \\ \dfrac{b(p_i)}{a(p_i)} - 1, & if\ a(p_i) > b(p_i) \end{cases}$$

For each topic, we computed the average $sw(p_i)$ over all data of a topic as a measure of how tightly grouped all the predicates in the topic are. Thus the average $sw(p_i)$ over all predicates of the entire dataset is a measure of how appropriately the predicates have been clustered.

The average $sw(p_i)$ over all predicates of each topic was computed at each level. For example, at level 1, K= 2 was computed using the SW. Furthermore, after partitioning into two topics, the silhouette widths, 0.89 (for 20 predicates) and 0.71 (for 43 predicates) are computed for each topic. At level 2, for the left topic, K= 5 and for the right topic, K = 2 were computed, respectively. After clustering, silhouette widths, 0.52 (for 4 predicates) and 0.7 (for 6 predicates), 0.59 (for 3 predicates), 0.92 (for 4 predicates), and 0.38 (for 3 predicates) and two silhouette widths, 0.76 (for 35 predicates) and 0.66 (for 8 predicates) were computed for each topic. At level 3, one of the topics were partitioned into two (K= 2). Two silhouette widths, 0.77 (for 20 predicates) and 0.65 (for 15 predicates) were computed for each topic. If there are too many or too few topics, as may occur when a poor choice of *k* in each level is used in the hierarchical K-means algorithm, some of the topics will typically display much narrower silhouettes than the rest. Thus silhouette averages are used to determine the number of topics within a dataset. We increased the likelihood of the silhouette (α = 0.5) being maximized at the correct number of topics by re-scaling the data using feature weights that are topic specific.

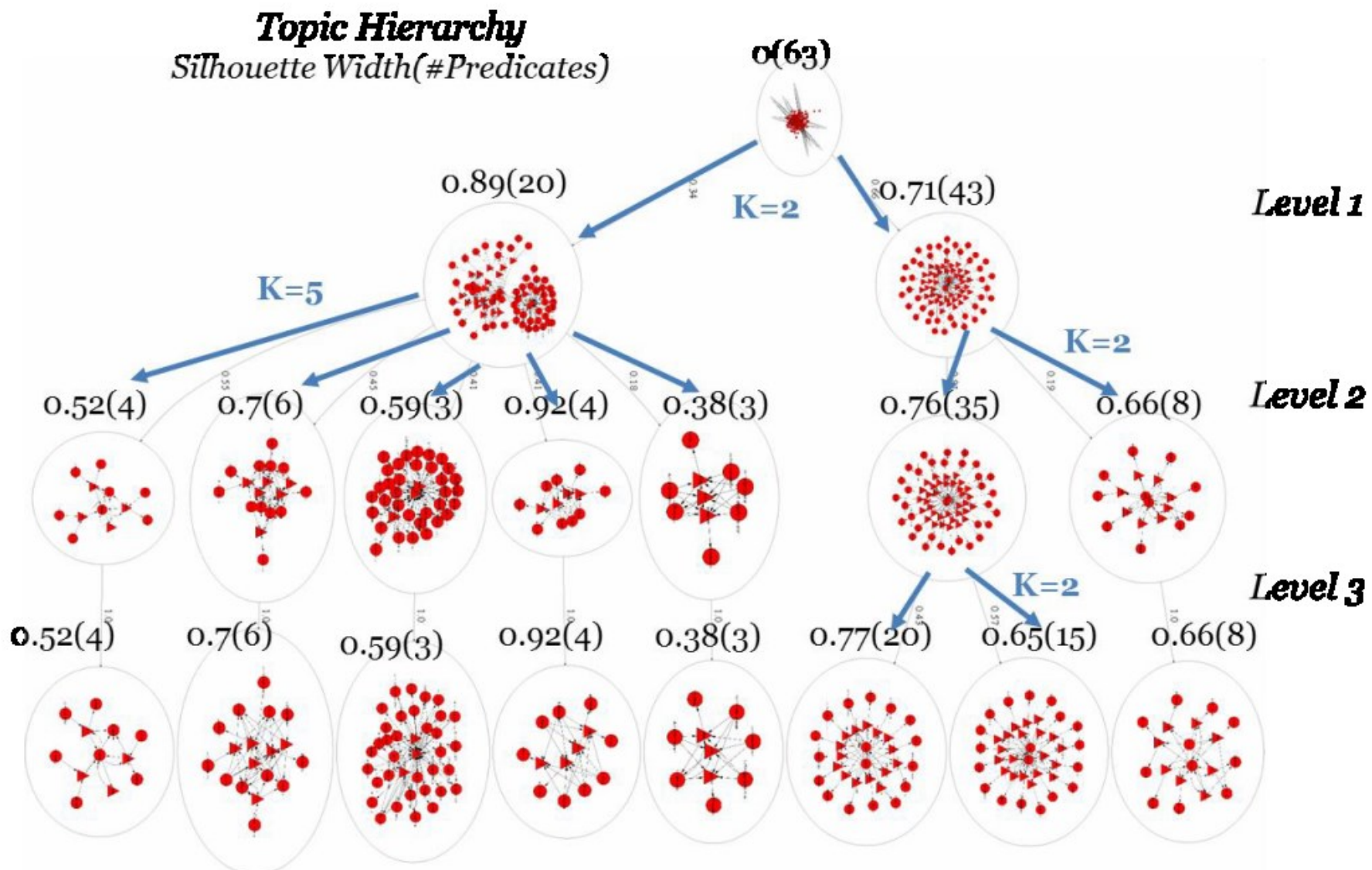

**Figure 4** Silhouette Width and Number of Topics in Topic Hierarchy

In the HPKM, a topic of interest was further clustered into K subtopics (the optimal K subtopics) using a heuristic algorithm, *Neighborhood Silhouette Width* (NSW). NSW is similar to the silhouette method that validates the consistency checking by examining how well each predicate fits some uniformity criterion in its cluster, whereas *Neighborhood Silhouette Width* (NSW) is the average of the weighted SW for the (neighbored) topics at a specific level that have the same parents. The *Neighborhood Silhouette Width* (NSW) is computed by the sum of the multiplication of silhouette width and the number of predicates in a particular topic, $\boldsymbol{NumP(T_i)}$, divided by the total number of predicates in the neighboring topics. The optimal *k* for a topic $\boldsymbol{T_l}$ at level *l* will be determined based on the highest *Neighborhood Silhouette Width* $\boldsymbol{nsw(T_l)}$

$$\boldsymbol{nsw(T_l) = \frac{\sum_{i=1}^{k} sw(T_i) * NumP(T_i))}{\sum_{i=1}^{k} NumP(T_i))}}$$

For example, as shown in Figure 4, for the given $\boldsymbol{sw(T_{1_1})}$ is 0.89 and $\boldsymbol{NumP(T_{1_1})}$ is 20 and $\boldsymbol{sw(T_{1_2})}$ is 0.71 and $\boldsymbol{NumP(T_{1_1})}$ is 43, the first level's *Neighborhood Silhouette Width* $\boldsymbol{nsw(T_{1_1})}$ is computed as follows $\boldsymbol{nsw(T_{1_1}) = \frac{(0.89*20+0.71*43)}{20+43} = 0.77}$. Therefore, at level 1, the highest NSW value is 0.77 and the optimal K is determined as 2. Similarly, the second level's *Neighborhood Silhouette Width* $\boldsymbol{nsw(T_{2_1})}$ is computed as followed: $\boldsymbol{nsw(T_{2_1}) = \frac{(0.52*4+0.7*6+0.59*3+0.92*4+0.38*3)}{4+6+3+4+3} = 0.64}$. and $\boldsymbol{nsw(T_{2_2}) = \frac{(0.76*35+0.66*8)}{35+8} = 0.74}$. Therefore, the highest NSW for $\boldsymbol{T_{2_1}}$ **and** $\boldsymbol{T_{2_2}}$ at level 2 is $(\boldsymbol{T_{2_1}})$ =0.64, $(\boldsymbol{T_{2_1}})$ =0.74 and the optimal K is determined as 5 and 2, respectively.

According to the optimal *k* determined by $\boldsymbol{nsw(T_l)}$, the level of the hierarchy that can represent topics at multiple tasks will be constructed at different levels until there is no further change in the hierarchy. If the silhouette width of each

Algorithm 1 Hierarchical Predicate-based K-Means Clustering (HPKM)

```
// P is an n × n predicate similarity matrix, n is the number of predicates in ontologies
// δ is the threshold of silhouette width C_ij
Input: P, δ
// a hierarchy with a set of clusters C_ij the j^th cluster at i^th level
Output: C = {C_11, C_12 ... C_ij}
i=1
repeat
    // n is the number of input predicates, m is the number of predicates of cluster j at level i,
    // optimal k (k ≤ m ≤ n) from m predicates of the cluster at level i (C_i) using nsw(c_i) function
    sw_1 = nsw(c_i)                    // compute Neighborhood silhouette width
```

$k$= OptimalK($C_i, sw_1$) // fine the optimal $k$ based on $nsw(C_i)$<br>
$Change1 = false$<br>
**If** (k > 1) then<br>
$Change1 = true$<br>
**for** $j = 1$ *to* $k$<br>
$\mu_{ij}$= $RM(p_{j1}, p_{j2} \dots p_{jm})$ // random mean for predicates in $C_{ij}$(cluster $j$ at level $i$)<br>
**end**<br>
**foreach** $p_{ij} \in P_i$ $do$<br>
$\mu_{ij}$ = $Argmin(p_{ij}, \mu_{ij})$ $j \in \{1 \dots k\}$<br>
**end**<br>
$Change2 = false$<br>
$sw = 0$<br>
**repeat**<br>
**foreach** $\mu_{ij} \in U_i$ $do$<br>
*UpdateCluster*($\mu_{ij}$)<br>
**end**<br>
**foreach** $p_{ij} \in P_i$ $do$<br>
$NCen = Argmin(p_{ij}, \mu_{ij})$ $j \in \{1 \dots k\}$<br>
**if** $NCen \neq \mu_{ij}$ then<br>
$\mu_{ij} = NCen$<br>
$C_{ij} = C_{ij} \cup p_{ij}$<br>
$changed2 = true$<br>
**endif**<br>
$sw_2$ = SilhouetteWidth ($C_{ij}$) // compute silhouette width<br>
**while** $Changed2 == true$<br>
**while** $Change1 == true$ and $sw_2 \geq \delta$

topic is lower than the threshold α, the clustering will be terminated. Thus, the maximum overall average silhouette width will be taken as the optimal clustering algorithms for the topic hierarchy. In this way, we can achieve the HPKM objective of maximizing intra-cluster similarities and minimizing inter-cluster similarities. The algorithm of Hierarchical Predicate-based K-Means clustering (HPKM) is given as shown in Algorithm 1.

### D. *Topic Ranking in Topic Hierarchy*

To characterize each topic in the hierarchy by an integrated rank, we computed the average value of the following five classifications: i) *Top 20 Predicates*, ii) *Top 20 Concepts*, iii) *Similarity*, iv) *Silhouette Width*, and v) *Density*. The first two rankings measure how popular they are relative to the rest of predicates and concepts. In determining the rankings for *Top 20 Predicates* and *Top 20 Concepts*, the weight of a predicate or concept that occurs in ontologies is simply proportional to the term frequency (about 20% and 30% were considered, respectively). For *Top 20 Concepts*, as there may be some duplicates among topics, the duplicates are eliminated before deciding the ranking. *Similarity* and *Silhouette Width* are measures for local (intra-relation of topics) and global (inter-relation of topics) similarity, respectively. Both measurements seem to be equally important in reflecting the importance of topics in a topic hierarchy. *Similarity* measurement was specified in Section II.B and silhouette width in Section II.C.

In order to measure the *Density*, we used network concepts, such as in-degree and out-degree of concepts (C) and predicates (P) in an ontology; in our model, C and P are vertexes, whereas V and the links between C and P are edges, E in a graph. For a predicate (P), the number of incoming edges (E) adjacent to a concept (C) is called the in-degree of the predicate (P) and the number of outgoing edges adjacent to a concept (C) is its out-degree of the predicate (P). The density (D) is computed as a ratio of the number of edges |E| to the number of possible edges between nodes ($|\boldsymbol{V}| = |\boldsymbol{P}| + |\boldsymbol{C}|$) as follows: $\boldsymbol{D} = \frac{\mathbf{2}|\boldsymbol{E}|}{(|\boldsymbol{C}|+|\boldsymbol{P}|)(|\boldsymbol{C}|-\mathbf{1})(|\boldsymbol{P}|-\mathbf{1})}$. The results of the topic ranking were used in query generation and query processing.

### E. *Query Generation*

From the HPKM, a topic hierarchy is generated. The Query Generation algorithm will start crawling the leaf nodes (the topics at the bottom level) in a given topic hierarchy and generate a *query* that is a part of a particular topic *TG* (a RDF graph) in the topic hierarchy. The algorithm will crawl the topic graph *TG t*o generate a query graph *QG*; *QG* is a subset of the *TG*. Many variation of queries can be generated from this process. The query generation algorithm automatically generates queries by traversing topic graphs. The topic graph has three predicates, namely *drug* from the Sider domain (in pink), *affected-organism* from the DrugBank domain (in red) and *x-pubchem-substance* from the Pharmacogenomics Knowledge Base (PharmGKB)[20] (in green).

We generated a query by traversing the predicate that has the highest rank δ (the highest sum of the in-degree and out-degree of the predicate) and traverse its neighbors level-by-level (Breadth-first Search) in the descending order of the similarity in the SM computed by the PNP algorithm. For this traversal, we consider the neighbors whose similarity scores are higher than threshold $\beta$. In this example, we started with the best predicate *drug* and then visit its neighbors whose similarity scores are higher than the threshold ($\beta = 0.2$) in a descending order. For example, *drug*, its nearest neighbor, *x-pubchem-substance* with the similarity score 0.5, thus we expand *drug* with an additional predicate*, x-pubchem-substance*. And then *drug's* next nearest neighbor is *affected-organism* with the similarity score 0.1. Since the similarity score is less than the threshold $\beta$, (i.e., $0.1 \leq 0.2$), we terminated the navigation. The algorithm runs until there is no more neighboring predicates to be considered. The generated query includes triples with two predicates, *drug* and *x-pubchem-substance,* and their subject variables (*?E* and *?D*) and object variables *(?D* and *?R*). The type of variable ?E is known as Drug Effect, ?D as Drug, and ?R as PharmGKB Resource. This can be converted to a triplet form such as <?D typeof Drug>. As shown in Figure 5, an example of the automatically was presented to demonstrate the generation of a SPARQL query[21] for a given topic graph.

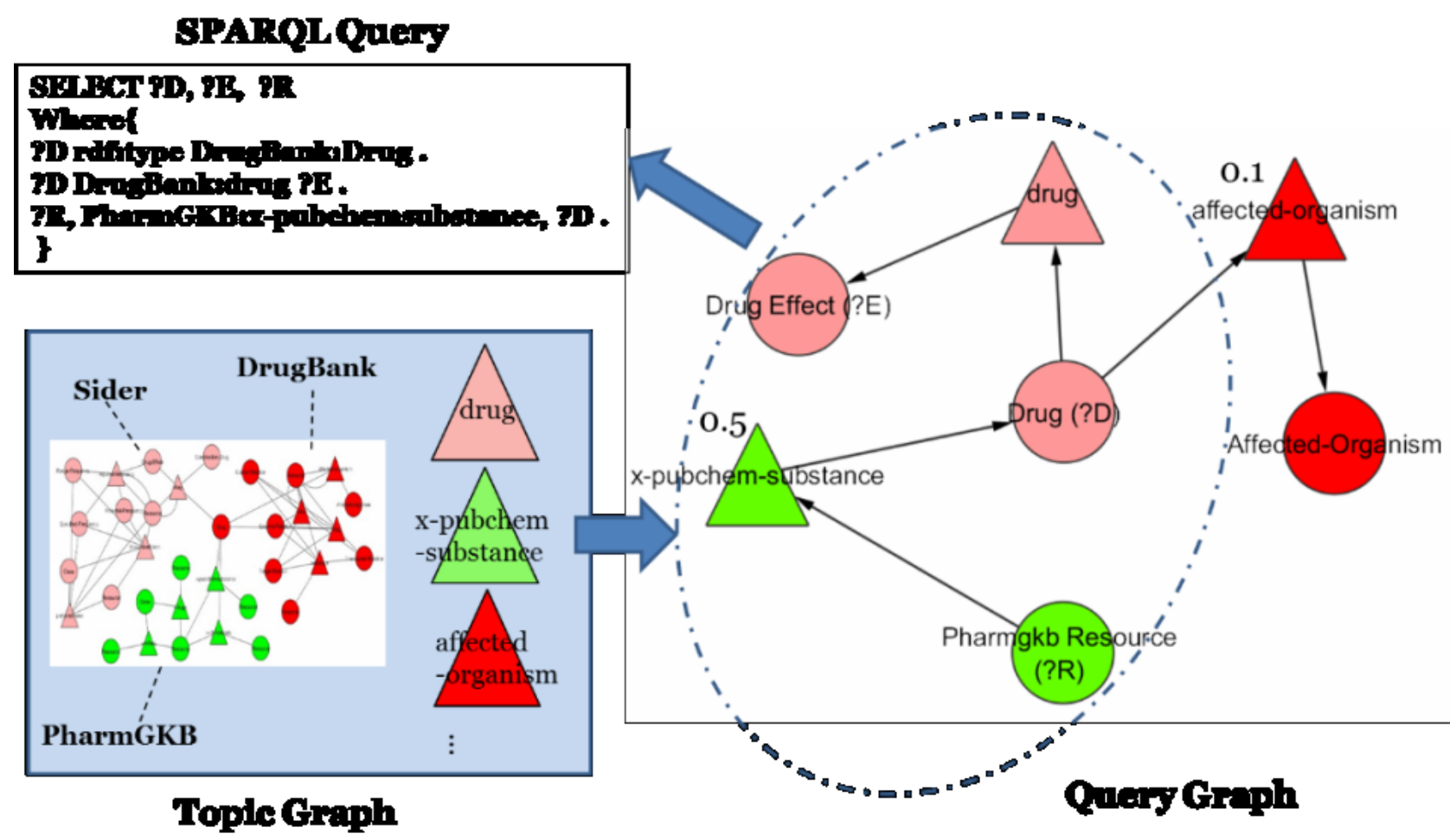

**Figure 5** Automatic Query Generation

## III. IMPLEMENTATION

The *MedTQ* system was implemented using Java in Eclipse Juno Integrated Development Environment[22]. Apache Jena API[23] was used to parse OWL/RDF datasets and retrieve triple information. We used R computing environment[24] for our experimental validation. We implemented a software plugin for query and schema graph visualization using CytoScape 3.0.2[25]. In addition, we have built a SPARQL query endpoint on a single machine that is hosted by the UMKC Distributed Intelligent Computing (UDIC) lab. (Figure 6) The OPEN LINK Virtuoso server version 6.1.3[26] was installed and the five Bio2RDF datasets (Bio2RDF ClinicalTrial[27], Bio2RDF DrugBank[28], Bio2RDF OMIM[29], Bio2RDF PharmGKB[30], and Bio2RDF Sider[31]) were imported into the graph domain http://Bio2RDF.com#.

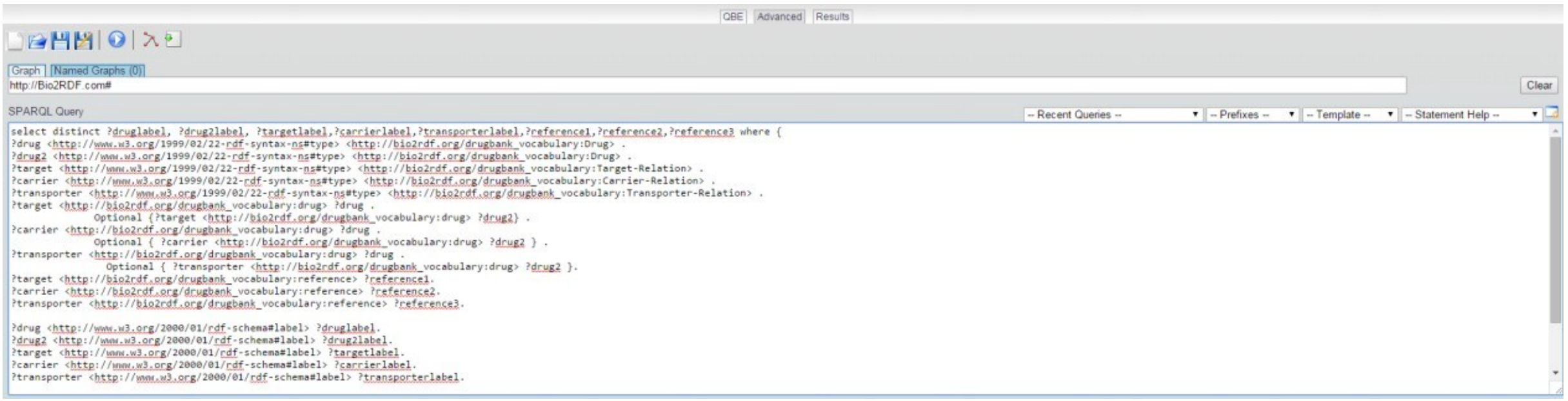

**Figure 6** SPARQL Endpoint

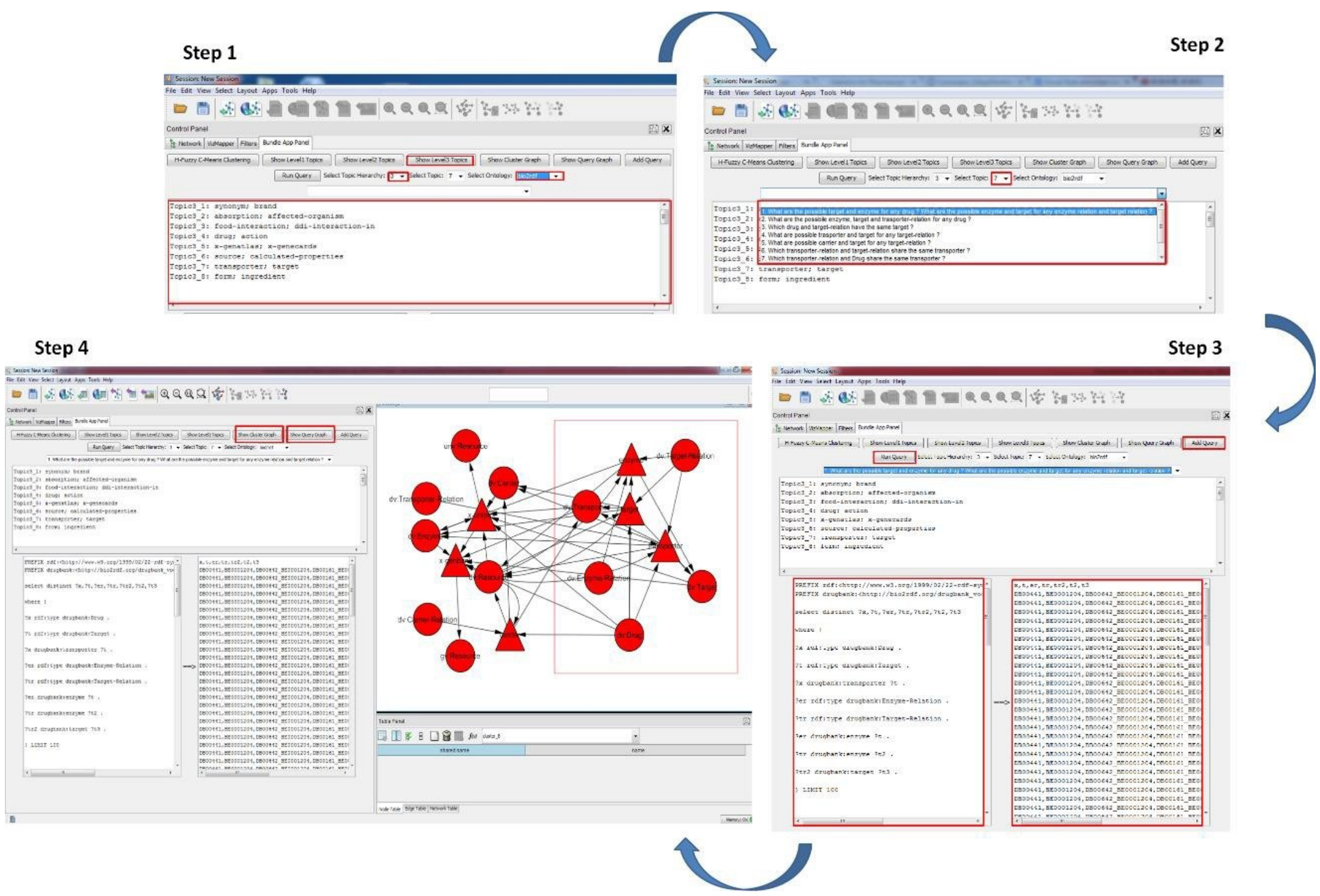

**Figure 7** MedTQ Interactive Query Tool

The *MedTQ* tool can be used for browsing the generated topics and performing interactive design and processing of queries. As shown in Figure 7, step 1 shows the list of topics for a given ontology (DrugBank). Step 2 shows the list of NLP questions for a selected topic (Topic 7). Step 3 shows the automatically generated SPARQL query and the query results. Step 4 shows the topic and query graphs for the selected query.

The steps for the query generation and processing using *MedTQ* tool are explained as follows:
**Step1**: A user first selects a dataset (e.g., DrugBank) to be analyzed, then choose an algorithm to generate a topic hierarchy (e.g., three level hierarchy). A clustering algorithm (e.g., Hierarchical K-Means Clustering) button is selected for the construction of a topic hierarchy (DrugBank). Topics generated from the topic hierarchy construction are listed in the top left box. In this example, the eight topics are shown with the detailed description including a list of the highest ranked predicates and their concepts (with high in-degree/out-degree).
**Step2**: The user selects a topic (e.g., 7$^{th}$ topic) to view, then this allows users to explore top ten natural language queries automatically generated by the proposed query generation algorithm.
**Step3**: A query can be selected and modified through the interactive query editor based on the topics or predicates shown in Step 2. Once the design of a query is complete, the corresponding natural language query expressions and the corresponding SPARQL query will be generated.
**Step4**: After choosing the natural language query expressions (e.g., *what are the enzyme, target and transporter-relation of a drug?*), the *add query* button can be clicked to select its corresponding SPARQL query into the bottom left box.

**Step5**: When the *query button* is clicked, the SPARQL query will be executed and the query output will be shown in the bottom right box.
**Step6**: When the *show query cluster* button is clicked, the corresponding cluster graph will be displayed on the canvas in the right panel. Moreover, by clicking the *show query graph* button, the relevant concepts and predicates in the SPARQL query will also be highlighted.

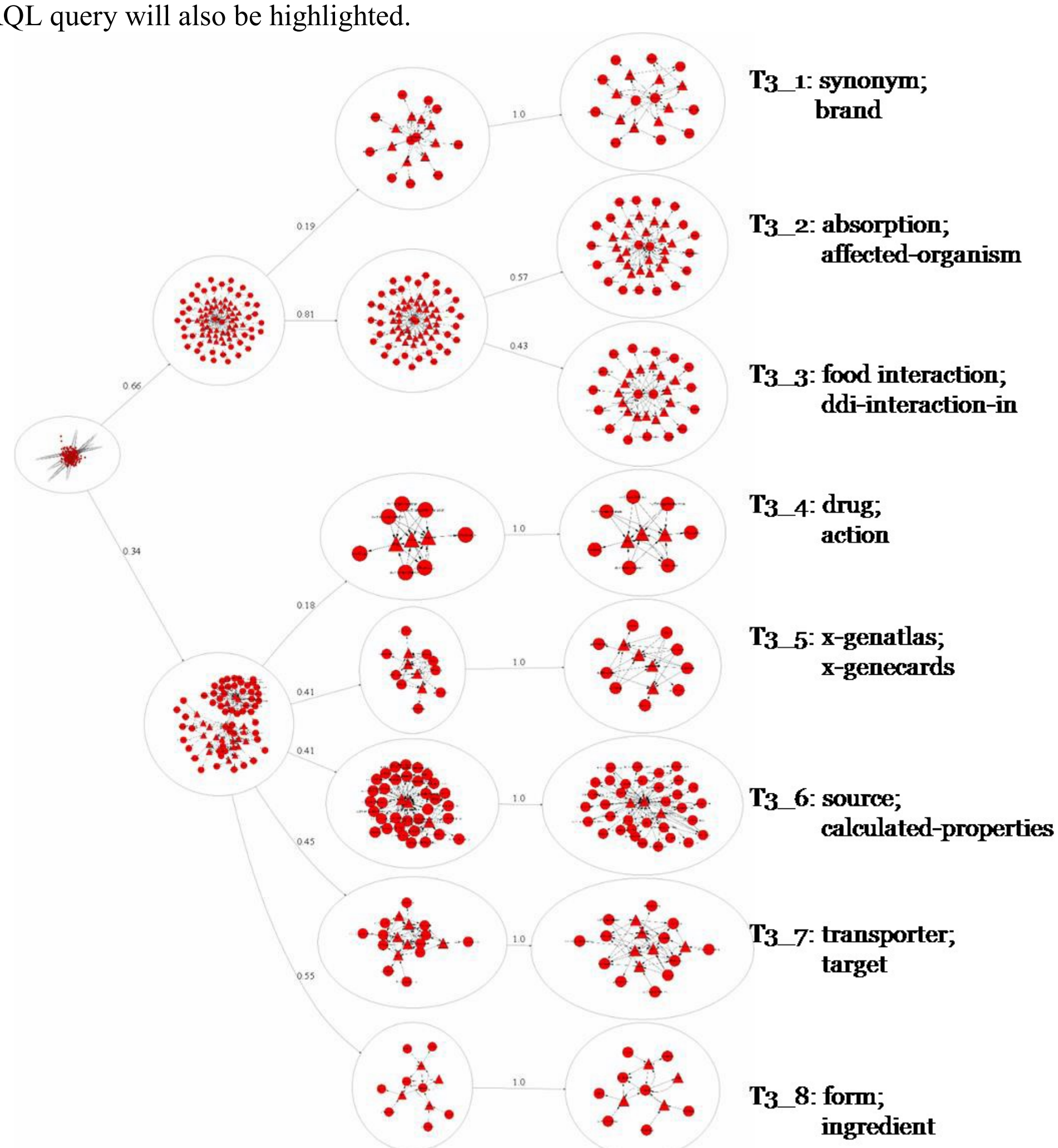

**Figure 8** DrugBank Topic Hierarchy

## IV. RESULTS

In this paper, we conducted a case study with a biomedical ontology (DrugBank). DrugBank is a public database with drug information on properties, structure, and biology of small molecule and biotech drugs. Important topics are drugs' *targets*, *enzymes*, *transporters*, and *carriers*. These may switch roles depending on the drug to which they bind so that some drugs specifically target transporters i.e., a transporter can also be the target. It is a key resource for bioinformatics and cheminformatics research. Based on DrugBank database and other life science data, Bio2RDF project[32] normalized them into a uniform format in a distributed network to support biomedical knowledge translation and discovery.

### A. *Topic Hierarchy Generated using HPKM Approach*

In this case study, we demonstrated the details of knowledge discovery as well as query generation in the proposed framework. We are particularly interested in generating interesting queries using the proposed PNP model and HPKM algorithms. In addition, the experiments have been conducted to validate the correctness of our approach. Table 1 shows the details of the DrugBank Ontology. In this case study, the unique concepts (C) of DrugBank ontology, excluding the duplicates, are considered. Only the domain specific predicates (P) excluding built-in predicates were considered. The number of edges in the graph (|E|) was computed as the sum of in-degree and out-degree. The overall density was computed based on the vertices (P+C) and the edges (E).

**Table 1.** DrugBank Ontology

| Features | Num | Features | Num |
|---|---|---|---|
| # Total Concepts | 116 | Sum of Indegree and Outdegree (\|E\|) | 519 |
| #Unique Concepts in DrugBank (C) | 93 | #Triples | 737 |
| # Total Predicates | 68 | # Domain Specific Triplets (T) | 401 |
| #Unique Domain Specific Predicate (P) | 63 | Density (D) | 0.043 |

The base URL of predicates is http://bio2rdf.org/drugbank_vocabulary. However, the concepts are from 25 different domains as shown in Table 2. Interestingly, all predicates are from the same domain and that gives us a good basis for linking concepts together either from same or different domains. This is one of the reasons we proposed a predicate-oriented approach. The concepts' domain URLs and their short notations are shown in Table 2.

**Table 2.** Short Notation for DrugBank and Related Domains

| prefix | Domain URL | prefix | Domain URL |
|---|---|---|---|
| ahv: | http://bio2rdf.org/ahfs_vocabulary: | kv: | http://bio2rdf.org/kegg_vocabulary: |
| av: | http://bio2rdf.org/atc_vocabulary: | owl: | http://www.w3.org/2002/07/owl# |
| bv: | http://bio2rdf.org/bindingdb_vocabulary: | pcv: | http://bio2rdf.org/pubchem.compound_vocabulary: |
| cv: | http://bio2rdf.org/chemspider_vocabulary: | pdv: | http://bio2rdf.org/pdb_vocabulary: |
| dv: | http://bio2rdf.org/drugbank_vocabulary: | psv: | http://bio2rdf.org/pubchem.substance_vocabulary: |
| dpv: | http://bio2rdf.org/dpd_vocabulary: | pv: | http://bio2rdf.org/pubmed_vocabulary: |
| gv: | http://bio2rdf.org/genbank_vocabulary: | uv: | http://bio2rdf.org/uspto_vocabulary: |
| gav: | http://bio2rdf.org/genatlas_vocabulary: | chv: | http://bio2rdf.org/chebi_vocabulary: |
| gcv: | http://bio2rdf.org/genecards_vocabulary: | nv: | http://bio2rdf.org/ndc_vocabulary: |
| giv: | http://bio2rdf.org/gi_vocabulary: | phv: | http://bio2rdf.org/pharmgkb_vocabulary: |
| gtv: | http://bio2rdf.org/gtp_vocabulary: | unv: | http://bio2rdf.org/uniprot_vocabulary: |
| hv: | http://bio2rdf.org/hgnc_vocabulary: | wv: | http://bio2rdf.org/wikipedia_vocabulary: |
| iv: | http://bio2rdf.org/iuphar_vocabulary: | | |

From the HPKM algorithm for each domain ontology, the *topic hierarchy* was generated. A topic hierarchy was generated for a single domain ontology, *DrugBank*. The topic hierarchy for *DrugBank* has the number of topics <2:7:8> with 2 topics at the first level, 7 topics at the second level, and 8 topics at the third level. K-means clustering was performed in a top-down manner until the average of clusters' silhouette width is higher than a certain threshold (> 0.5). The number on each edge in the topic hierarchy represents the percentage of predicates that the upper level topic graph contributes to the lower level graph. For example, for the two topics in the first level of *DrugBank*, 66% of predicates of the *DrugBank* ontology are contributed to Topic 1 (T1_1) while 34% to Topic 2 (T1_2). The contribution rate is ranged between 0 and 1. Interestingly, predicates are unique to their topic graph, however, some concepts in a topic may appear in more than one topics. Moreover, for each topic at $3^{rd}$ level, top 2 ranked predicates (computed based on in-degree/out-degree) were selected as a representative term for each topic.

B. *Top Predicates and Concepts of DrugBank*

Table 3 shows the ranks for Top 20 predicates and Top 20 concepts that were computed in terms of the sum of their in-degree and out-degree. These predicates and concepts are shown in terms of Predicate Rank (PR), Predicates, Predicate IO (PIO) and Predicate Topic ID (PIO), corresponding Concepts, Concept Rank (CR), and together with the description of predicates specified by DrugBank. From this list, many top predicates are from Topic 3_6, Topic 3_7, and Topic 3_1. Many top concepts are from Topic 3_7, Topic 3_2, and Topic 3_3. The prefix dv: of these concepts indicates the domain http://bio2rdf.org/drugbank_vocabulary. Some of the Top 20 Concepts were not directly mapped with the predicates in the Top 20 Predicates. These concepts are *dv:Enzyme-Relation*, *dv:Target-Relation*, *dv:Carrier-Relation*, *dv:LogP*, *dv:LogS*, *dv:Molecular-Formula*, *dv:Molecular-Weight*, *dv:Transporter-Relation*, *dv:Water-Solubility*, *dv:Bioavailability*, *dv:Boiling-Point*, *dv:Caco2-Permeability*. These results show that the predicates rankings are not always the same with the concept rankings.

**Table 3.** Top 20 Predicates and Concepts in DrugBank Ontology

| PR | Predicates | PIO | TID | CR | Concepts | Description |
|---|---|---|---|---|---|---|
| 1 | source | 66 | T3_6 | 46 | dv:Source | Data sources indicate the source of the information present in each drug properties. |
| 2 | calculated-properties | 56 | T3_6 | | N/A | The calculation from the structure for the drug. |
| 3 | experimental-properties | 28 | T3_6 | | N/A | A medicinal product may be approved for use in one disease or condition but still be considered experimental for other diseases or conditions. |
| 4 | transporter | 17 | T3_7 | 7 | dv:Transporter | A membrane bound protein which shuttles ions, small molecules or macromolecules across membranes, into cells or out of cells. |
| 5 | target | 16 | T3_7 | 6 | dv:Target | A protein, macromolecule, nucleic acid, or small molecule to which a given drug binds, resulting in an alteration of the normal function of the bound molecule and a desirable therapeutic effect. |
| 6 | drug | 14 | T3_4 | 2 | dv:Drug | A drug is a chemical substance that, when ingested, has a high biological response to quantity ratio compared to regular foods. |
| 7 | enzyme | 13 | T3_7 | 5 | dv:Enzyme | A protein which catalyzes chemical reactions involving a given drug (substrate). |
| 8 | carrier | 12 | T3_7 | 4 | dv:Carrier | Drug carriers may be used in drug design to increase the effectiveness of drug delivery to the target sites of pharmacological actions. |
| 9 | action | 11 | T3_4 | | N/A | The drugs that enter the human tend to stimulate certain receptors, ion channels, act on enzymes or transporter proteins. As a result, they cause the human body to react in a specific way. |
| 10 | synonym | 10 | T3_1 | 21 | dv:Synonym | Alternate names of the drug |
| 11 | brand | 9 | T3_1 | 51 | dv:Brand | Brand names from different manufacturers |
| 12 | category | 8 | T3_1 | 21 | dv:Category | Category of drugs |
| 13 | form | 8 | T3_8 | | N/A | Dosage forms are essentially pharmaceutical drug products in the form in which they are marketed for use. |
| 14 | ingredient | 8 | T3_8 | | N/A | An ingredient is a substance that forms part of a mixture |
| 15 | x-genbank | 7 | T3_7 | | N/A | Link between DrugBank and GenBank |
| 16 | x-uniprot | 7 | T3_7 | | N/A | Link between DrugBank and UniProt |
| 17 | manufacturer | 6 | T3_1 | 51 | dv:Manufacturer | Companies known to manufacture the given drug |
| 18 | mixture | 6 | T3_1 | 45 | dv:Mixture | A mixture refers to the physical combination of two or more substances on which the identities are retained and are mixed in the form of solutions, suspensions, and colloids. |
| 19 | toxicity | 6 | T3_1 | 51 | dv:Toxicity | Lethal dose (LD50) values from test animals, description of side effects and toxic effects seen in humans |
| 20 | absorption | 6 | T3_2 | 51 | dv:Absorption | Description of how much of the drug or how readily the drug is taken up by the body |

Table 4 shows the duplicated concepts among topics. The total number of instances is 40 and the number of duplicates is 23. *dv:Resource* and *dv:Drug* appear in almost all the topics. According to this analysis, the sets of the topic groups {T3_1 and T3_8}, {T3_4 and T3_7}, and {T3_5 and T3_7} are similar. However, these are quite different from the outcomes from the predicated-oriented clustering algorithm.

**Table 4.** Duplicated Concepts and their Topic ID in DrugBank Ontology

| Concepts | Freq | Topics | Concepts | Freq | Topics |
|---|---|---|---|---|---|
| dv:Resource | 8 | T3_1, T3_2, T3_3, T3_4, T3_5, T3_6, T3_7, T3_8 | dv:Transporter-Relation | 2 | T3_4, T3_7 |
| dv:Drug | 6 | T3_1, T3_2, T3_3, T3_4, T3_6, T3_7 | dv:Enzyme-Relation | 2 | T3_4, T3_7 |
| uv:Resource | 2 | T3_1, T3_8 | dv:Carrier | 2 | T3_5, T3_7 |
| dv:Mixture | 2 | T3_1, T3_8 | dv:Target | 2 | T3_5, T3_7 |
| dv:Patent | 2 | T3_1, T3_8 | dv:Enzyme | 2 | T3_5, T3_7 |
| dv:Pharmaceutical | 2 | T3_2, T3_8 | dv:Transporter | 2 | T3_5, T3_7 |
| dv:Carrier-Relation | 2 | T3_4, T3_7 | uv:Resource | 2 | T3_1, T3_8 |
| dv:Target-Relation | 2 | T3_4, T3_7 | | | |

C. *Validation for Hierarchical K-Means Clustering*

An experiment has been conducted to find an optimal number of the clusters using the four different clustering algorithms, K-means[17], Clustering Large Application (Clara)[33], Partition Around Medoids (Pam)[34], and Hierarchical Clustering[16]. The results of the optimal K validation algorithm presented in Section II.C based on the clustering outcomes by the four different algorithms. As shown in Figure 9, Clara, Pam and Hierarchical clustering algorithms are not a good approach to find an optimal cluster number since they show a relative stable silhouette width for varying the number of clusters. The proposed HPKM algorithm determines the most significant number of clusters at each level such as K = 2 with SW = 0.77 at level 1 and K = 5 with SW = 0.64 and K=2 with SW = 0.74 at level 2 and K = 2 with SW = 0.72 at level 3. The HPKM algorithm was validated and compared against other algorithms in terms of the cluster number and the silhouette width.

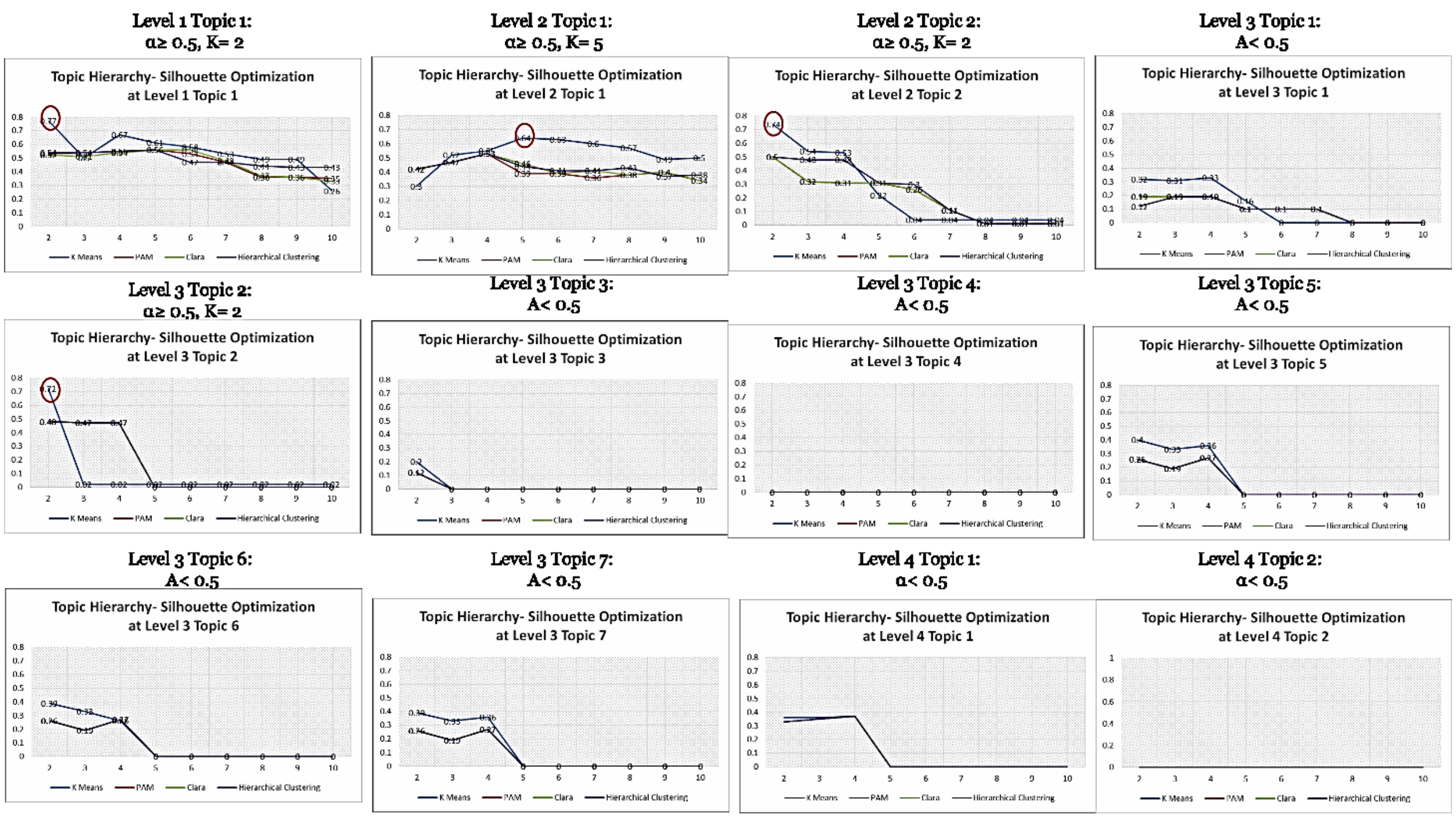

**Figure 9** Optimal K Branching Factors using Multiple Clustering Techniques

D. *Results for Topic Generation*

For the DrugBank ontology, we've considered 63 concepts, 116 predicates. The relevance scales of five different rankings and an overall ranking are evaluated. As shown in Figure 10, overall rank was computed in terms of the following 5 criteria: i) *Top 20 Concepts*, ii) *Top 20 Predicates*, iii) *Similarity*, iv) *Silhouette Width*, v) *Density*. Eight topics ranked from best to worst as follows: Topic 3_4, Topic 3_7, Topic 3_6, Topic 3_2, Topic 3_1, Topic 3_3, Topic 3_8, and Topic 3_5. Specifically, Topic 3_4 shows the best ranking for all three criteria such as *Similarity*, *Silhouette Width* and *Density*. However, Topic 3_7's *Top 20 Concept Ranking*, *Top 20 Property Ranking*, and *Similarity Ranking* are relatively good. From the ranking results, we have observed that the proposed ranking system correctly captured Topic 3_4 and Topic 3_7 as the core topics of DrugBank. Topic 3_5 was ranked the worst among the eight topics. Since Topic 3_5 is a connector topic whose predicates are mainly used to connect DrugBank with other domains. It is relatively less important from a single domain (DrugBank) perspective. However, Topic 3_5 would be very useful from a cross domain perspective.

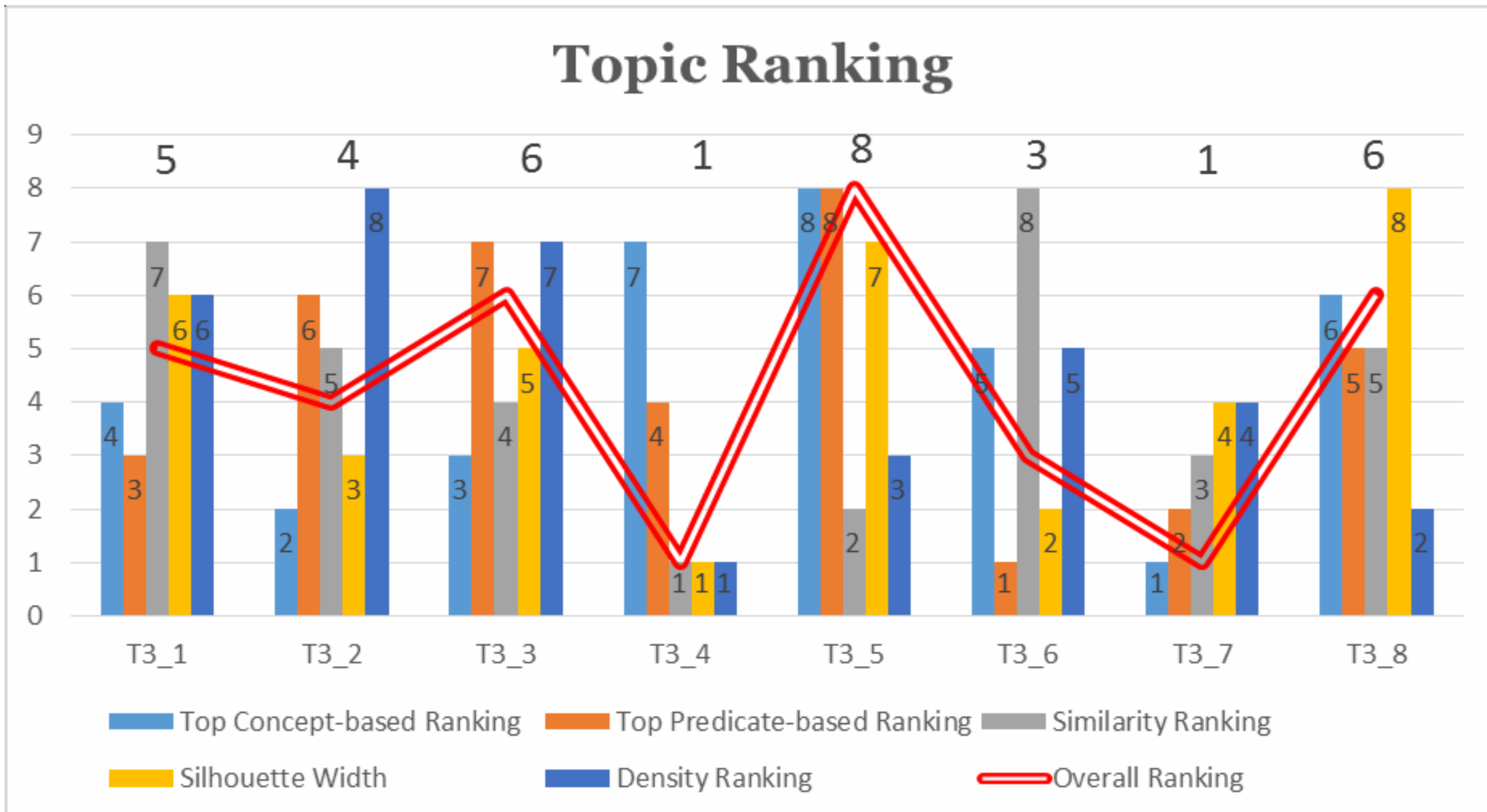

**Figure 10** Topic Rankings for DrugBank

E. *Results for Topic Discovery & Query Generation*

We now show the four best topic graphs at level 3 of the DrugBank topic hierarchy as follows: Topic 3_4, Topic 3_7, Topic 3_6, and Topic 3_2. In addition, the automatically generated query and query results of each topic are also shown.

**Rank 1: Topic 3_4 (T3_4):** T3_4's overall rank is 1$^{st}$ among 8 topics. This topic graph consists of 6 predicates and 12 concepts with 72 in-degree and out-degree. As shown in Figure 11, among 12 concepts, five concepts (*dv:Resource*; *dv:Drug*, *dv:Carrier*, *dv:Enzyme*, *dv:Target*) are ranked among *Top 20 Concepts* and all 6 predicates of this topic are ranked among *Top 20 Predicates*. In particular, there are two groups of predicates; one is with four predicates such as *transporter*, *target*, *enzyme*, carrier with concepts *dv:Target-Relation*, *dv:Target-Relation*, *dv:Enzyme-Relation*, *dv:Carrier-Relation*, respectively. Another group of predicates such as *x-genbank* and *x-uniprot* is a connector predicate group that is mainly used to connect between internal concepts (e.g., *dv:Drug*, *dv:Enzyme*) and external concepts (e.g., *gv:Resource*, *unv:Resource*). Specifically, T3_4 shows very high rankings for *Similarity*, *Silhouette Width*, and *Density* while showing a relatively low ranking for *Top 20 Concepts*. In this graph, concepts are represented as a circle, predicates as a triangle, and links as an arrow. In addition, the dark red items are the predicates and concepts mentioned in Query-1.

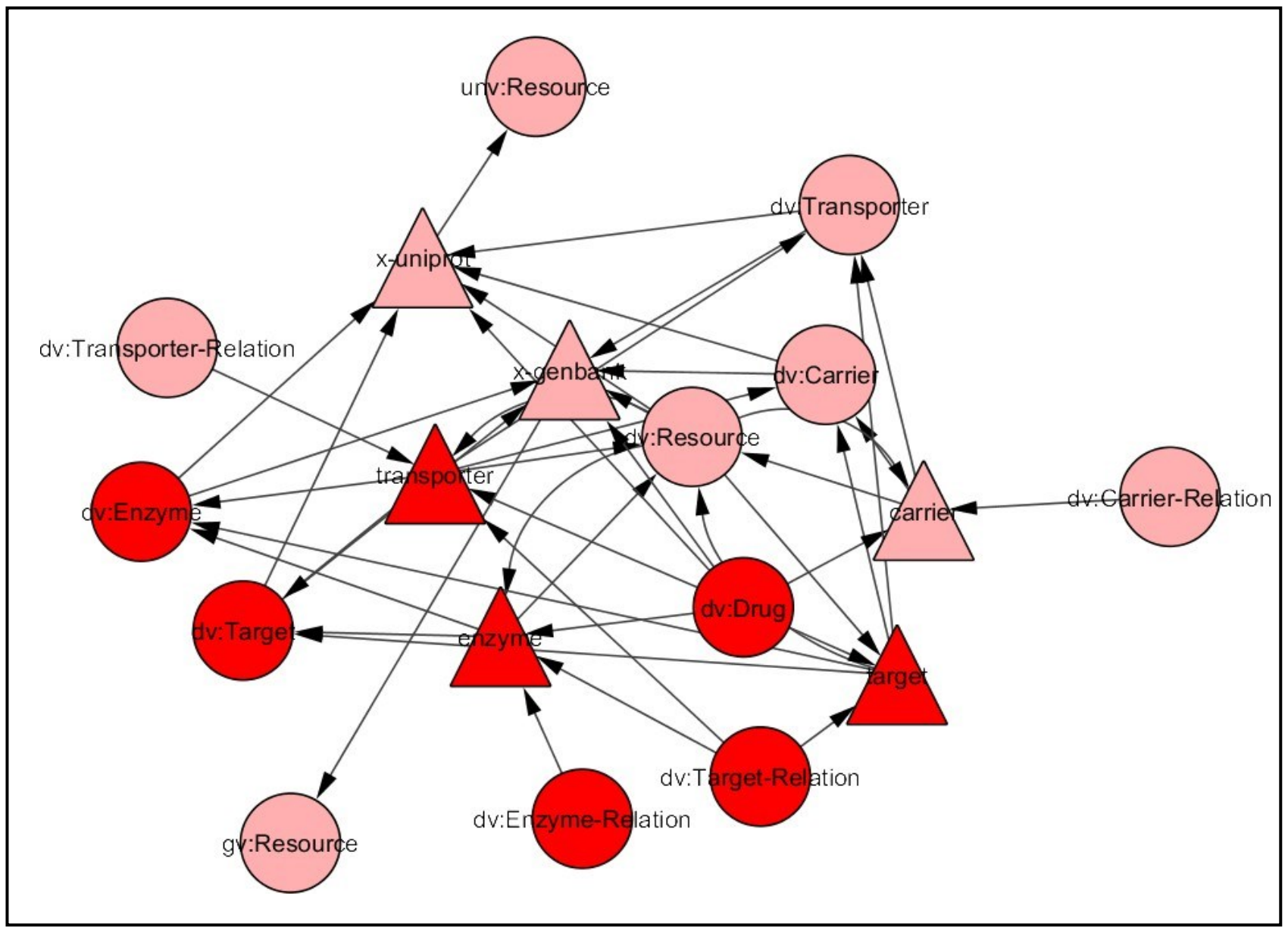

**Figure 11** Topic_3_4 (Topic 4 at Level 3) Graph in DrugBank

**Query-1:** The query is automatically generated from Topic 3_4 (one of the top ranked topics) by our query generation algorithm. This query allows users to find the most relevant drugs in terms of their *target, enzyme, enzyme relation, and target relation*. The SPARQL format of Query-1 was automatically generated by the Query Generation algorithm (described in Section II-E) considering the top predicates and their concepts (described in Section IV-B) below.

***Q1. For any two drugs which share the same target and transporter enzyme, what are all the possible drugs, enzyme, target, enzyme relations, target relations?***

```
select distinct ?druglabel, ?targetlabel, ?erlabel, ?trlabel, ?drug2label, ?enzymelabel where {
?drug <http://www.w3.org/1999/02/22-rdf-syntax-ns#type> <http://bio2rdf.org/drugbank_vocabulary:Drug> .
?target <http://www.w3.org/1999/02/22-rdf-syntax-ns#type> <http://bio2rdf.org/drugbank_vocabulary:Target> .
?drug <http://bio2rdf.org/drugbank_vocabulary:target> ?target .
?drug <http://bio2rdf.org/drugbank_vocabulary:transporter> ?enzyme .
?er <http://www.w3.org/1999/02/22-rdf-syntax-ns#type> <http://bio2rdf.org/drugbank_vocabulary:Enzyme-Relation> .
?tr <http://www.w3.org/1999/02/22-rdf-syntax-ns#type> <http://bio2rdf.org/drugbank_vocabulary:Target-Relation> .
?er <http://bio2rdf.org/drugbank_vocabulary:enzyme> ?enzyme .
?tr <http://bio2rdf.org/drugbank_vocabulary:enzyme> ?target.
?drug2 <http://bio2rdf.org/drugbank_vocabulary:target> ?target .
?drug2 <http://bio2rdf.org/drugbank_vocabulary:transporter> ?enzyme .
?drug <http://www.w3.org/2000/01/rdf-schema#label> ?druglabel.
?target <http://www.w3.org/2000/01/rdf-schema#label> ?targetlabel.
?er <http://www.w3.org/2000/01/rdf-schema#label> ?erlabel.
?tr <http://www.w3.org/2000/01/rdf-schema#label> ?trlabel.
?drug2 <http://www.w3.org/2000/01/rdf-schema#label> ?drug2label.
?enzyme <http://www.w3.org/2000/01/rdf-schema#label> ?enzymelabel. }
```

The Query-1 results include Gemcitabine, Fluorouracil, Ribavirin as the relevant drugs, Thymidylate synthase and Adenosine kinaseas as the target and Equilibrative nucleoside transporter 1 as the enzyme. Specifically, as shown in Figure 12, partial outputs of Query-1 include information on drug target and transporter enzyme of some drugs (e.g., Gemcitabine [drugbank:DB00441], Fluorouracil [drugbank:DB00544], Ribavirin [drugbank:DB00811]) and their enzyme relations, target relations for any drugs that share the same target and transporter enzyme.

| druglabel | targetlabel | erlabel | trlabel | drug2label | enzymelabel |
|---|---|---|---|---|---|
| Gemcitabine [drugbank:DB00441] | Thymidylate synthase [drugbank:BE0000324] | drugbank:DB00642 to drugbank:BE0001204 relation [drugbank_resource:DB00642_BE0001204] | drugbank:DB00544 to drugbank:BE0000324 relation [drugbank_resource:DB00544_BE0000324] | Gemcitabine [drugbank:DB00441] | Equilibrative nucleoside transporter 1 [drugbank:BE0001204] |
| Fluorouracil [drugbank:DB00544] | Thymidylate synthase [drugbank:BE0000324] | drugbank:DB00642 to drugbank:BE0001204 relation [drugbank_resource:DB00642_BE0001204] | drugbank:DB00544 to drugbank:BE0000324 relation [drugbank_resource:DB00544_BE0000324] | Gemcitabine [drugbank:DB00441] | Equilibrative nucleoside transporter 1 [drugbank:BE0001204] |
| Gemcitabine [drugbank:DB00441] | Thymidylate synthase [drugbank:BE0000324] | drugbank:DB00642 to drugbank:BE0001204 relation [drugbank_resource:DB00642_BE0001204] | drugbank:DB00544 to drugbank:BE0000324 relation [drugbank_resource:DB00544_BE0000324] | Fluorouracil [drugbank:DB00544] | Equilibrative nucleoside transporter 1 [drugbank:BE0001204] |
| Fluorouracil [drugbank:DB00544] | Thymidylate synthase [drugbank:BE0000324] | drugbank:DB00642 to drugbank:BE0001204 relation [drugbank_resource:DB00642_BE0001204] | drugbank:DB00544 to drugbank:BE0000324 relation [drugbank_resource:DB00544_BE0000324] | Fluorouracil [drugbank:DB00544] | Equilibrative nucleoside transporter 1 [drugbank:BE0001204] |
| Ribavirin [drugbank:DB00811] | Adenosine kinase [drugbank:BE0003540] | drugbank:DB00642 to drugbank:BE0001204 relation [drugbank_resource:DB00642_BE0001204] | drugbank:DB00811 to drugbank:BE0003540 relation [drugbank_resource:DB00811_BE0003540] | Ribavirin [drugbank:DB00811] | Equilibrative nucleoside transporter 1 [drugbank:BE0001204] |

**Figure 12** Partial Results of Query-1 in DrugBank

**Rank 2: Topic 3_7 (T3_7):** T3_7's overall rank is 2nd among 8 topics (together with T3_4). As shown in Figure 13, this topic graph is composed of 7 concepts represented as a circle and 3 predicates as a triangle with 31 in-degree and out-degree. In T3_7, three predicates, *drug*, *action*, *reference*, whose in-degree and out-degree are 14, 11, and 6, respectively, are all nicely connected with 7 concepts. The predicates *drug* and *action* are ranked at 6th, 9th and many of the concepts in this topic are ranked among Top 20 Concepts. For T3_7, the rankings for *Top 20 Concepts*, *Top 20 Predicates*, *Similarity*, *Silhouette Width*, and *Density* are very good. In this graph, concepts are represented as a circle, predicates as a triangle, and links as an arrow. In addition, the dark red items like drug and reference are the predicates mentioned in Query-2.

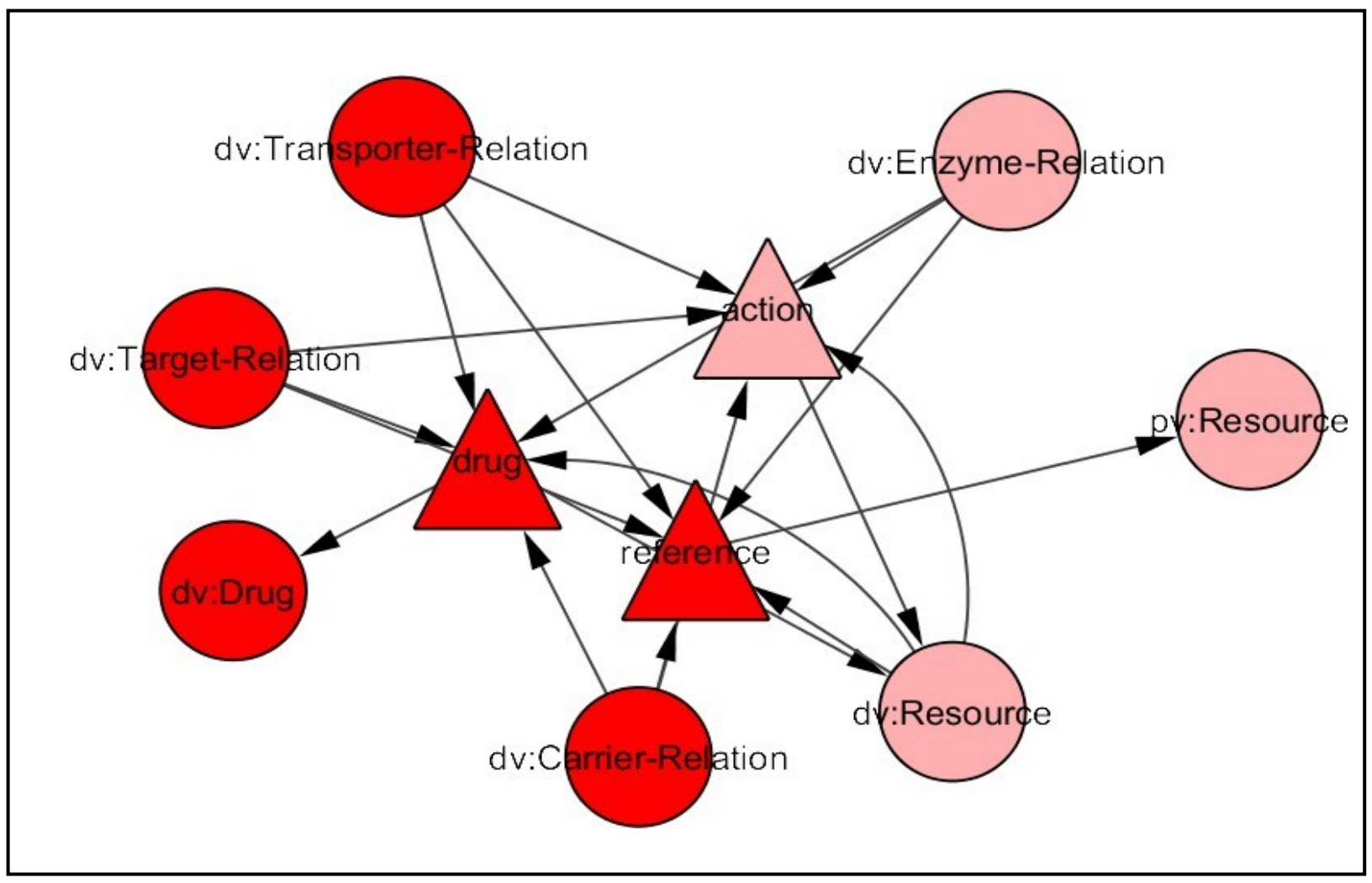

**Figure 13** Topic 3_7 (Topic 7 at Level 3) Graph in DrugBank

**Query-2:** The query graph was automatically generated from Topic 3_4 (one of the top ranked topics) to depict the query information. This query allows users to find the relevant drugs that have *common Target-Relation, Carrier-Relation and Transporter-Relation* and also provide their *PubMed references* for relations of target, transporter, and carrier with these drugs. The SPARQL format of Query-2 was automatically generated by the Query Generation algorithm (described in Section II-E) considering the top predicates and their concepts (described in Section IV-B) below.

***Q2. For any two drugs which share the common target-relation, carrier-relation and transporter-relation, what are all the possible combinations? What are the pubmed references for these target-relations, carrier-relation and transporter-relation?***

```
select distinct ?druglabel, ?drug2label, ?targetlabel,?carrierlabel,?transporterlabel,?reference1,?reference2,?reference3 where
{
?drug <http://www.w3.org/1999/02/22-rdf-syntax-ns#type> <http://bio2rdf.org/drugbank_vocabulary:Drug> .
?drug2 <http://www.w3.org/1999/02/22-rdf-syntax-ns#type> <http://bio2rdf.org/drugbank_vocabulary:Drug> .
?target <http://www.w3.org/1999/02/22-rdf-syntax-ns#type> <http://bio2rdf.org/drugbank_vocabulary:Target-Relation> .
?carrier <http://www.w3.org/1999/02/22-rdf-syntax-ns#type> <http://bio2rdf.org/drugbank_vocabulary:Carrier-Relation> .
?transporter <http://www.w3.org/1999/02/22-rdf-syntax-ns#type> <http://bio2rdf.org/drugbank_vocabulary:Transporter-
Relation> .
?target <http://bio2rdf.org/drugbank_vocabulary:drug> ?drug .
        Optional {?target <http://bio2rdf.org/drugbank_vocabulary:drug> ?drug2} .
?carrier <http://bio2rdf.org/drugbank_vocabulary:drug> ?drug .
        Optional { ?carrier <http://bio2rdf.org/drugbank_vocabulary:drug> ?drug2 } .
?transporter <http://bio2rdf.org/drugbank_vocabulary:drug> ?drug .
         Optional { ?transporter <http://bio2rdf.org/drugbank_vocabulary:drug> ?drug2 }.
?target <http://bio2rdf.org/drugbank_vocabulary:reference> ?reference1.
?carrier <http://bio2rdf.org/drugbank_vocabulary:reference> ?reference2.
?transporter <http://bio2rdf.org/drugbank_vocabulary:reference> ?reference3.

?drug <http://www.w3.org/2000/01/rdf-schema#label> ?druglabel.
?drug2 <http://www.w3.org/2000/01/rdf-schema#label> ?drug2label.
?target <http://www.w3.org/2000/01/rdf-schema#label> ?targetlabel.
?carrier <http://www.w3.org/2000/01/rdf-schema#label> ?carrierlabel.
?transporter <http://www.w3.org/2000/01/rdf-schema#label> ?transporterlabel.
}
```

As shown in Figure 14, partial results from the query include information about some drugs such as *Phenytoin* (DrugBank:DB00252), *Lepirudin* (DrugBank:DB00001) and *Deferasirox* (DrugBank:DB01609). The results show all the possible combinations and their PubMed references for *Target-relation*, *Carrier-relation,* and *Transporter-relation* of drugs sharing the common information of *Target-relation*, *Carrier-relation,* and *Transporter-relation*.

| druglabel | drug2label | targetlabel | carrierlabel | transporterlabel | reference1 | reference2 | reference3 |
|---|---|---|---|---|---|---|---|
| Phenytoin [drugbank:DB00252] | Lepirudin [drugbank:DB00001] | drugbank:DB00252 to drugbank:BE0000141 relation [drugbank_resource:DB00252_BE0000141] | drugbank:DB00252 to drugbank:BE0000530 relation [drugbank_resource:DB00252_BE0000530] | drugbank:DB00252 to drugbank:BE0001032 relation [drugbank_resource:DB00252_BE0001032] | pubmed:15805193 | pubmed:15282104 | pubmed:17045309 |
| Phenytoin [drugbank:DB00252] | Lepirudin [drugbank:DB00001] | drugbank:DB00252 to drugbank:BE0000141 relation [drugbank_resource:DB00252_BE0000141] | drugbank:DB00252 to drugbank:BE0000530 relation [drugbank_resource:DB00252_BE0000530] | drugbank:DB00252 to drugbank:BE0001032 relation [drugbank_resource:DB00252_BE0001032] | pubmed:17001291 | pubmed:15282104 | pubmed:17045309 |
| Phenytoin [drugbank:DB00252] | Lepirudin [drugbank:DB00001] | drugbank:DB00252 to drugbank:BE0000141 relation [drugbank_resource:DB00252_BE0000141] | drugbank:DB00252 to drugbank:BE0000530 relation [drugbank_resource:DB00252_BE0000530] | drugbank:DB00252 to drugbank:BE0001032 relation [drugbank_resource:DB00252_BE0001032] | pubmed:20298965 | pubmed:15282104 | pubmed:17045309 |
| Phenytoin [drugbank:DB00252] | Lepirudin [drugbank:DB00001] | drugbank:DB00252 to drugbank:BE0000141 relation [drugbank_resource:DB00252_BE0000141] | drugbank:DB00252 to drugbank:BE0000530 relation [drugbank_resource:DB00252_BE0000530] | drugbank:DB00252 to drugbank:BE0001032 relation [drugbank_resource:DB00252_BE0001032] | pubmed:15805193 | pubmed:16621742 | pubmed:17045309 |
| Phenytoin [drugbank:DB00252] | Lepirudin [drugbank:DB00001] | drugbank:DB00252 to drugbank:BE0000141 relation [drugbank_resource:DB00252_BE0000141] | drugbank:DB00252 to drugbank:BE0000530 relation [drugbank_resource:DB00252_BE0000530] | drugbank:DB00252 to drugbank:BE0001032 relation [drugbank_resource:DB00252_BE0001032] | pubmed:17001291 | pubmed:16621742 | pubmed:17045309 |
| Phenytoin [drugbank:DB00252] | Lepirudin [drugbank:DB00001] | drugbank:DB00252 to drugbank:BE0000141 relation [drugbank_resource:DB00252_BE0000141] | drugbank:DB00252 to drugbank:BE0000530 relation [drugbank_resource:DB00252_BE0000530] | drugbank:DB00252 to drugbank:BE0001032 relation [drugbank_resource:DB00252_BE0001032] | pubmed:20298965 | pubmed:16621742 | pubmed:17045309 |
| Phenytoin [drugbank:DB00252] | Deferasirox [drugbank:DB01609] | drugbank:DB00252 to drugbank:BE0000141 relation [drugbank_resource:DB00252_BE0000141] | drugbank:DB00252 to drugbank:BE0000530 relation [drugbank_resource:DB00252_BE0000530] | drugbank:DB00252 to drugbank:BE0001032 relation [drugbank_resource:DB00252_BE0001032] | pubmed:15805193 | pubmed:15282104 | pubmed:17045309 |
| Phenytoin [drugbank:DB00252] | Deferasirox [drugbank:DB01609] | drugbank:DB00252 to drugbank:BE0000141 relation [drugbank_resource:DB00252_BE0000141] | drugbank:DB00252 to drugbank:BE0000530 relation [drugbank_resource:DB00252_BE0000530] | drugbank:DB00252 to drugbank:BE0001032 relation [drugbank_resource:DB00252_BE0001032] | pubmed:17001291 | pubmed:15282104 | pubmed:17045309 |
| Phenytoin [drugbank:DB00252] | Deferasirox [drugbank:DB01609] | drugbank:DB00252 to drugbank:BE0000141 relation [drugbank_resource:DB00252_BE0000141] | drugbank:DB00252 to drugbank:BE0000530 relation [drugbank_resource:DB00252_BE0000530] | drugbank:DB00252 to drugbank:BE0001032 relation [drugbank_resource:DB00252_BE0001032] | pubmed:20298965 | pubmed:15282104 | pubmed:17045309 |
| Phenytoin [drugbank:DB00252] | Deferasirox [drugbank:DB01609] | drugbank:DB00252 to drugbank:BE0000141 relation [drugbank_resource:DB00252_BE0000141] | drugbank:DB00252 to drugbank:BE0000530 relation [drugbank_resource:DB00252_BE0000530] | drugbank:DB00252 to drugbank:BE0001032 relation [drugbank_resource:DB00252_BE0001032] | pubmed:15805193 | pubmed:16621742 | pubmed:17045309 |
| Phenytoin [drugbank:DB00252] | Deferasirox [drugbank:DB01609] | drugbank:DB00252 to drugbank:BE0000141 relation [drugbank_resource:DB00252_BE0000141] | drugbank:DB00252 to drugbank:BE0000530 relation [drugbank_resource:DB00252_BE0000530] | drugbank:DB00252 to drugbank:BE0001032 relation [drugbank_resource:DB00252_BE0001032] | pubmed:17001291 | pubmed:16621742 | pubmed:17045309 |
| Phenytoin [drugbank:DB00252] | Deferasirox [drugbank:DB01609] | drugbank:DB00252 to drugbank:BE0000141 relation [drugbank_resource:DB00252_BE0000141] | drugbank:DB00252 to drugbank:BE0000530 relation [drugbank_resource:DB00252_BE0000530] | drugbank:DB00252 to drugbank:BE0001032 relation [drugbank_resource:DB00252_BE0001032] | pubmed:20298965 | pubmed:16621742 | pubmed:17045309 |

**Figure 14** Results of Query-2 in DrugBank

**Rank 3: Topic 3_6 (T3_6)**: As shown in Figure 15, this topic consists of 3 predicates and 34 concepts with a high sum of in-degree and out-degree, 150. In particular, there are two subgraphs; one is with two predicates such as *source* and *calculated-properties* with concepts *dv:Boiling-Point* and *dv:Bioavailability*, respectively. Another predicate *experimental-properties* is connected with concepts such as *dv:Water-Solubility*. Specifically, T3_6 highly ranked in *Top 20 Predicates* and *Silhouette Width* while being lowly ranked in *Similarity*. This means each predicate has their own concepts while having the least common concepts with other predicates. Since the similarity ranking of this topic is low, the shared information is limited. Interestingly, this graph shows a connection pattern from *dv:experimental-properties* to *dv: source*. The overall rank is 3rd among the eight topics. In this graph, the dark red items like source and calculated-properties are the predicates mentioned in Query-3.

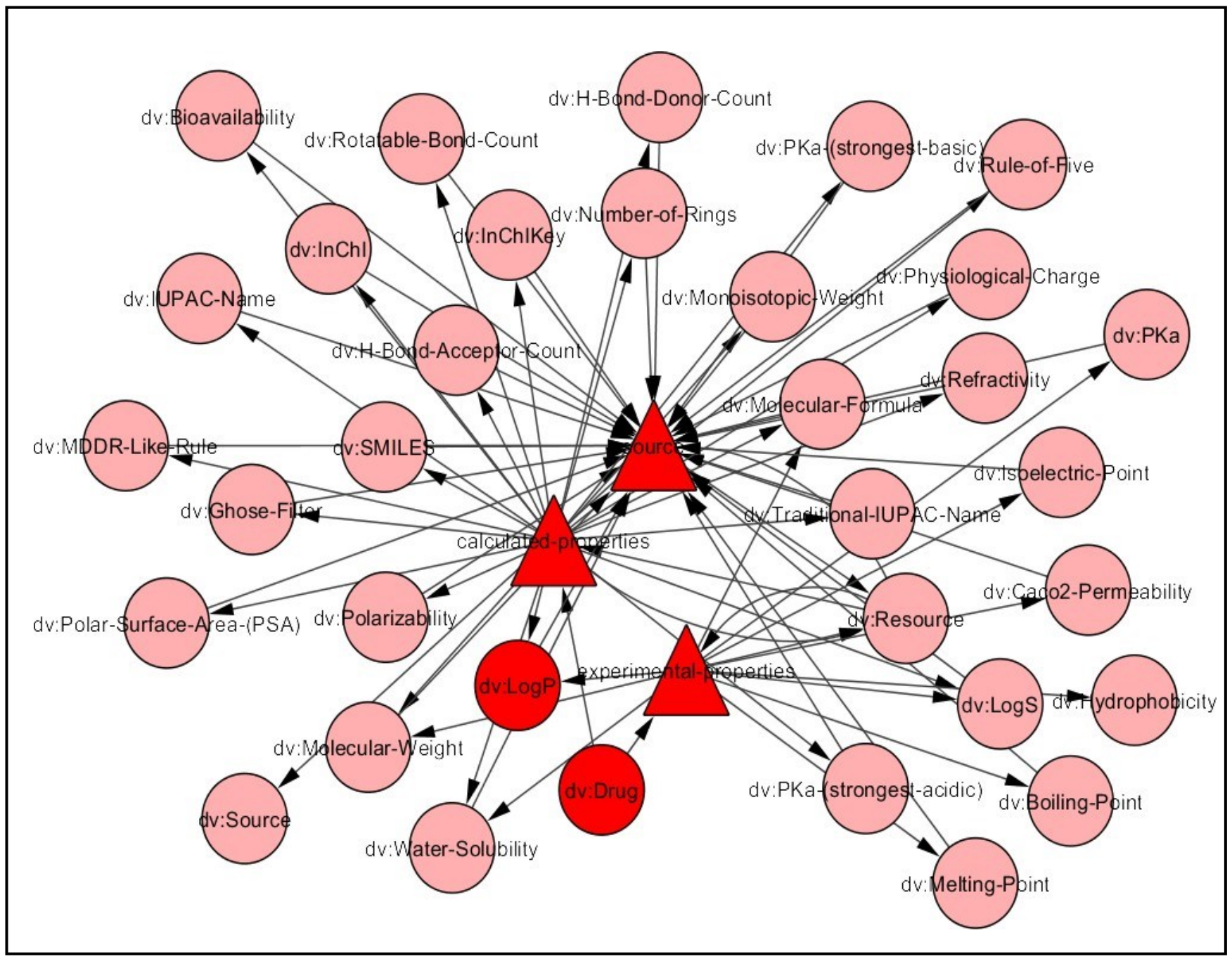

**Figure 15** Graph of Query-3 in DrugBank

**Query-3**. The query graph was automatically generated from Topic 3_6 to depict the query information. This query allows users to find drugs and *all their experimental properties and calculated properties which have LogP experimental properties (octanol-water partition coefficient)*. The SPARQL format of Query-3 was automatically generated by the Query Generation algorithm (described in Section II-E) considering the top predicates and their concepts (described in Section IV-B) below.

***Q3. For any drug, what are all its experimental properties and calculated properties which contain octanol-water partition coefficient?***

```
select distinct ?druglabel,?logp1label,?logp2label,?source1label,?source2label where {
?drug <http://www.w3.org/1999/02/22-rdf-syntax-ns#type> <http://bio2rdf.org/drugbank_vocabulary:Drug> .
?logp1 <http://www.w3.org/1999/02/22-rdf-syntax-ns#type> <http://bio2rdf.org/drugbank_vocabulary:LogP> .
?logp2 <http://www.w3.org/1999/02/22-rdf-syntax-ns#type> <http://bio2rdf.org/drugbank_vocabulary:LogP> .
?drug <http://bio2rdf.org/drugbank_vocabulary:experimental-properties> ?logp1 .
?drug <http://bio2rdf.org/drugbank_vocabulary:calculated-properties> ?logp2 .
?logp1 <http://bio2rdf.org/drugbank_vocabulary:source> ?source1 .
?logp2 <http://bio2rdf.org/drugbank_vocabulary:source> ?source2 .
?drug <http://www.w3.org/2000/01/rdf-schema#label> ?druglabel.
?logp1 <http://www.w3.org/2000/01/rdf-schema#label> ?logp1label.
?logp2 <http://www.w3.org/2000/01/rdf-schema#label> ?logp2label.
?source1 <http://www.w3.org/2000/01/rdf-schema#label> ?source1label.
?source2 <http://www.w3.org/2000/01/rdf-schema#label> ?source2label.
}
```

As the Query-3 results shown in Figure 16, the relevant drug and their experimental and calculated-properties are reported as {L-Histidine, logP: -3.32 from CHMELIK,J ET AL. (1991), logP: -3.1 from ALOGPS} and L-Phenylalanine, logP: -1.38 from AVDEEF,A (1997), logP: -1.4 from ALOGPS}. The partial query results on Query-3 include the information on some drugs (e.g., L-Histidine [drugbank:DB00117], L-Phenylalanine [drugbank:DB00120], L-Arginine [drugbank:DB00125]) and all their experimental properties and calculated properties that have *LogP experimental properties (octanol-water partition coefficient)*.

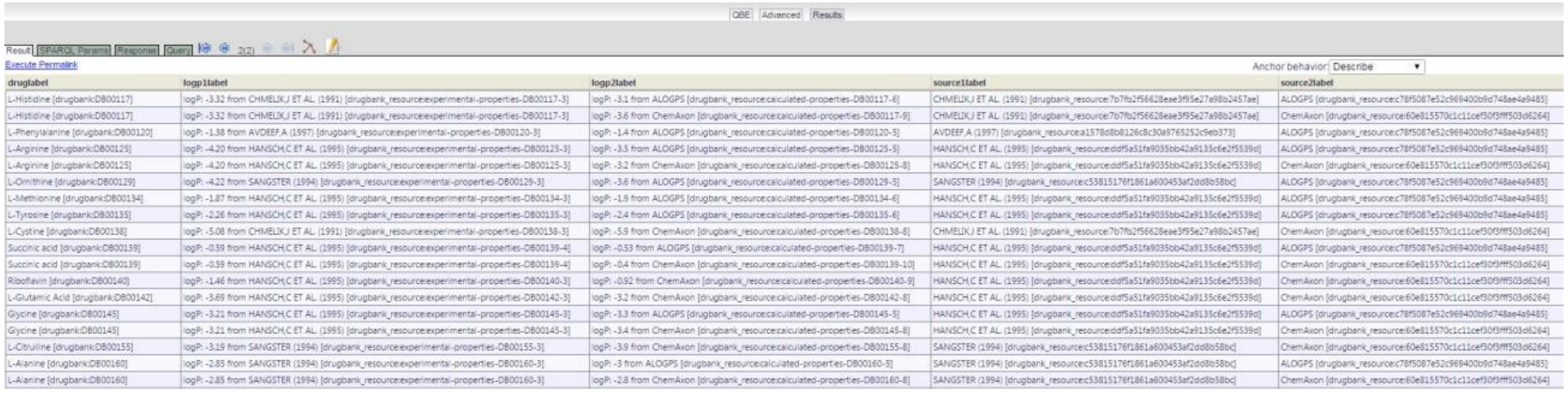

QBE Advanced Results

Result SPARQL Params Response Query 2(2)

Execute Permalink

Anchor behavior: Describe

| druglabel | logp1label | logp2label | source1label | source2label |
|---|---|---|---|---|
| L-Histidine [drugbank:DB00117] | logP: -3.32 from CHMELIK,J ET AL. (1991) [drugbank_resource:experimental-properties-DB00117-3] | logP: -3.1 from ALOGPS [drugbank_resource:calculated-properties-DB00117-6] | CHMELIK,J ET AL. (1991) [drugbank_resource:7b7fb2f56628eae3f95e27a98b2457ae] | ALOGPS [drugbank_resource:c78f5087e52c969400b9d748ae4a9485] |
| L-Histidine [drugbank:DB00117] | logP: -3.32 from CHMELIK,J ET AL. (1991) [drugbank_resource:experimental-properties-DB00117-3] | logP: -3.6 from ChemAxon [drugbank_resource:calculated-properties-DB00117-9] | CHMELIK,J ET AL. (1991) [drugbank_resource:7b7fb2f56628eae3f95e27a98b2457ae] | ChemAxon [drugbank_resource:60e815570c1c11cef30f3fff503d6264] |
| L-Phenylalanine [drugbank:DB00120] | logP: -1.38 from AVDEEF,A (1997) [drugbank_resource:experimental-properties-DB00120-3] | logP: -1.4 from ALOGPS [drugbank_resource:calculated-properties-DB00120-5] | AVDEEF,A (1997) [drugbank_resource:a1578d8b8126c8c30a9765252c9eb373] | ALOGPS [drugbank_resource:c78f5087e52c969400b9d748ae4a9485] |
| L-Arginine [drugbank:DB00125] | logP: -4.20 from HANSCH,C ET AL. (1995) [drugbank_resource:experimental-properties-DB00125-3] | logP: -3.5 from ALOGPS [drugbank_resource:calculated-properties-DB00125-5] | HANSCH,C ET AL. (1995) [drugbank_resource:ddf5a51fa9035bb42a9135c6e2f5539d] | ALOGPS [drugbank_resource:c78f5087e52c969400b9d748ae4a9485] |
| L-Arginine [drugbank:DB00125] | logP: -4.20 from HANSCH,C ET AL. (1995) [drugbank_resource:experimental-properties-DB00125-3] | logP: -3.2 from ChemAxon [drugbank_resource:calculated-properties-DB00125-8] | HANSCH,C ET AL. (1995) [drugbank_resource:ddf5a51fa9035bb42a9135c6e2f5539d] | ChemAxon [drugbank_resource:60e815570c1c11cef30f3fff503d6264] |
| L-Ornithine [drugbank:DB00129] | logP: -4.22 from SANGSTER (1994) [drugbank_resource:experimental-properties-DB00129-3] | logP: -3.6 from ALOGPS [drugbank_resource:calculated-properties-DB00129-5] | SANGSTER (1994) [drugbank_resource:c53815176f1861a600453af2dd8b58bc] | ALOGPS [drugbank_resource:c78f5087e52c969400b9d748ae4a9485] |
| L-Methionine [drugbank:DB00134] | logP: -1.87 from HANSCH,C ET AL. (1995) [drugbank_resource:experimental-properties-DB00134-3] | logP: -1.9 from ALOGPS [drugbank_resource:calculated-properties-DB00134-6] | HANSCH,C ET AL. (1995) [drugbank_resource:ddf5a51fa9035bb42a9135c6e2f5539d] | ALOGPS [drugbank_resource:c78f5087e52c969400b9d748ae4a9485] |
| L-Tyrosine [drugbank:DB00135] | logP: -2.26 from HANSCH,C ET AL. (1995) [drugbank_resource:experimental-properties-DB00135-3] | logP: -2.4 from ALOGPS [drugbank_resource:calculated-properties-DB00135-6] | HANSCH,C ET AL. (1995) [drugbank_resource:ddf5a51fa9035bb42a9135c6e2f5539d] | ALOGPS [drugbank_resource:c78f5087e52c969400b9d748ae4a9485] |
| L-Cystine [drugbank:DB00138] | logP: -5.08 from CHMELIK,J ET AL. (1991) [drugbank_resource:experimental-properties-DB00138-3] | logP: -5.9 from ChemAxon [drugbank_resource:calculated-properties-DB00138-8] | CHMELIK,J ET AL. (1991) [drugbank_resource:7b7fb2f56628eae3f95e27a98b2457ae] | ChemAxon [drugbank_resource:60e815570c1c11cef30f3fff503d6264] |
| Succinic acid [drugbank:DB00139] | logP: -0.59 from HANSCH,C ET AL. (1995) [drugbank_resource:experimental-properties-DB00139-4] | logP: -0.53 from ALOGPS [drugbank_resource:calculated-properties-DB00139-7] | HANSCH,C ET AL. (1995) [drugbank_resource:ddf5a51fa9035bb42a9135c6e2f5539d] | ALOGPS [drugbank_resource:c78f5087e52c969400b9d748ae4a9485] |
| Succinic acid [drugbank:DB00139] | logP: -0.59 from HANSCH,C ET AL. (1995) [drugbank_resource:experimental-properties-DB00139-4] | logP: -0.4 from ChemAxon [drugbank_resource:calculated-properties-DB00139-10] | HANSCH,C ET AL. (1995) [drugbank_resource:ddf5a51fa9035bb42a9135c6e2f5539d] | ChemAxon [drugbank_resource:60e815570c1c11cef30f3fff503d6264] |
| Riboflavin [drugbank:DB00140] | logP: -1.46 from HANSCH,C ET AL. (1995) [drugbank_resource:experimental-properties-DB00140-3] | logP: -0.92 from ChemAxon [drugbank_resource:calculated-properties-DB00140-9] | HANSCH,C ET AL. (1995) [drugbank_resource:ddf5a51fa9035bb42a9135c6e2f5539d] | ChemAxon [drugbank_resource:60e815570c1c11cef30f3fff503d6264] |
| L-Glutamic Acid [drugbank:DB00142] | logP: -3.69 from HANSCH,C ET AL. (1995) [drugbank_resource:experimental-properties-DB00142-3] | logP: -3.2 from ChemAxon [drugbank_resource:calculated-properties-DB00142-8] | HANSCH,C ET AL. (1995) [drugbank_resource:ddf5a51fa9035bb42a9135c6e2f5539d] | ChemAxon [drugbank_resource:60e815570c1c11cef30f3fff503d6264] |
| Glycine [drugbank:DB00145] | logP: -3.21 from HANSCH,C ET AL. (1995) [drugbank_resource:experimental-properties-DB00145-3] | logP: -3.3 from ALOGPS [drugbank_resource:calculated-properties-DB00145-5] | HANSCH,C ET AL. (1995) [drugbank_resource:ddf5a51fa9035bb42a9135c6e2f5539d] | ALOGPS [drugbank_resource:c78f5087e52c969400b9d748ae4a9485] |
| Glycine [drugbank:DB00145] | logP: -3.21 from HANSCH,C ET AL. (1995) [drugbank_resource:experimental-properties-DB00145-3] | logP: -3.4 from ChemAxon [drugbank_resource:calculated-properties-DB00145-8] | HANSCH,C ET AL. (1995) [drugbank_resource:ddf5a51fa9035bb42a9135c6e2f5539d] | ChemAxon [drugbank_resource:60e815570c1c11cef30f3fff503d6264] |
| L-Citrulline [drugbank:DB00155] | logP: -3.19 from SANGSTER (1994) [drugbank_resource:experimental-properties-DB00155-3] | logP: -3.9 from ChemAxon [drugbank_resource:calculated-properties-DB00155-8] | SANGSTER (1994) [drugbank_resource:c53815176f1861a600453af2dd8b58bc] | ChemAxon [drugbank_resource:60e815570c1c11cef30f3fff503d6264] |
| L-Alanine [drugbank:DB00160] | logP: -2.85 from SANGSTER (1994) [drugbank_resource:experimental-properties-DB00160-3] | logP: -3 from ALOGPS [drugbank_resource:calculated-properties-DB00160-5] | SANGSTER (1994) [drugbank_resource:c53815176f1861a600453af2dd8b58bc] | ALOGPS [drugbank_resource:c78f5087e52c969400b9d748ae4a9485] |
| L-Alanine [drugbank:DB00160] | logP: -2.85 from SANGSTER (1994) [drugbank_resource:experimental-properties-DB00160-3] | logP: -2.8 from ChemAxon [drugbank_resource:calculated-properties-DB00160-8] | SANGSTER (1994) [drugbank_resource:c53815176f1861a600453af2dd8b58bc] | ChemAxon [drugbank_resource:60e815570c1c11cef30f3fff503d6264] |

**Figure 16** Results of Query-3 in DrugBank

**Rank 4: Topic 3_2 (T3_2)**: As shown in Figure 17, unique pattern in T3_2 is two dominant concepts, *Resource* and *Drug*, whose in-degree and out-degree are 46 and 33, respectively, are fully connected to the remaining 17 concepts via 20 different predicates such as *absorption*, *protein-binding*. The rankings for *Top 20 Concepts* and *Silhouette Width* are relatively good while the rankings for *Top 20 Predicates* and *Density* are poor. The overall rank is 4th among the eight topics. In this graph, concepts are represented as a circle, predicates as a triangle, and links as an arrow. The dark red items like *abortion* and *product* are the predicates and *dv:Drug* and *dv:Pharmaceutical* are the concepts mentioned in Query-4.

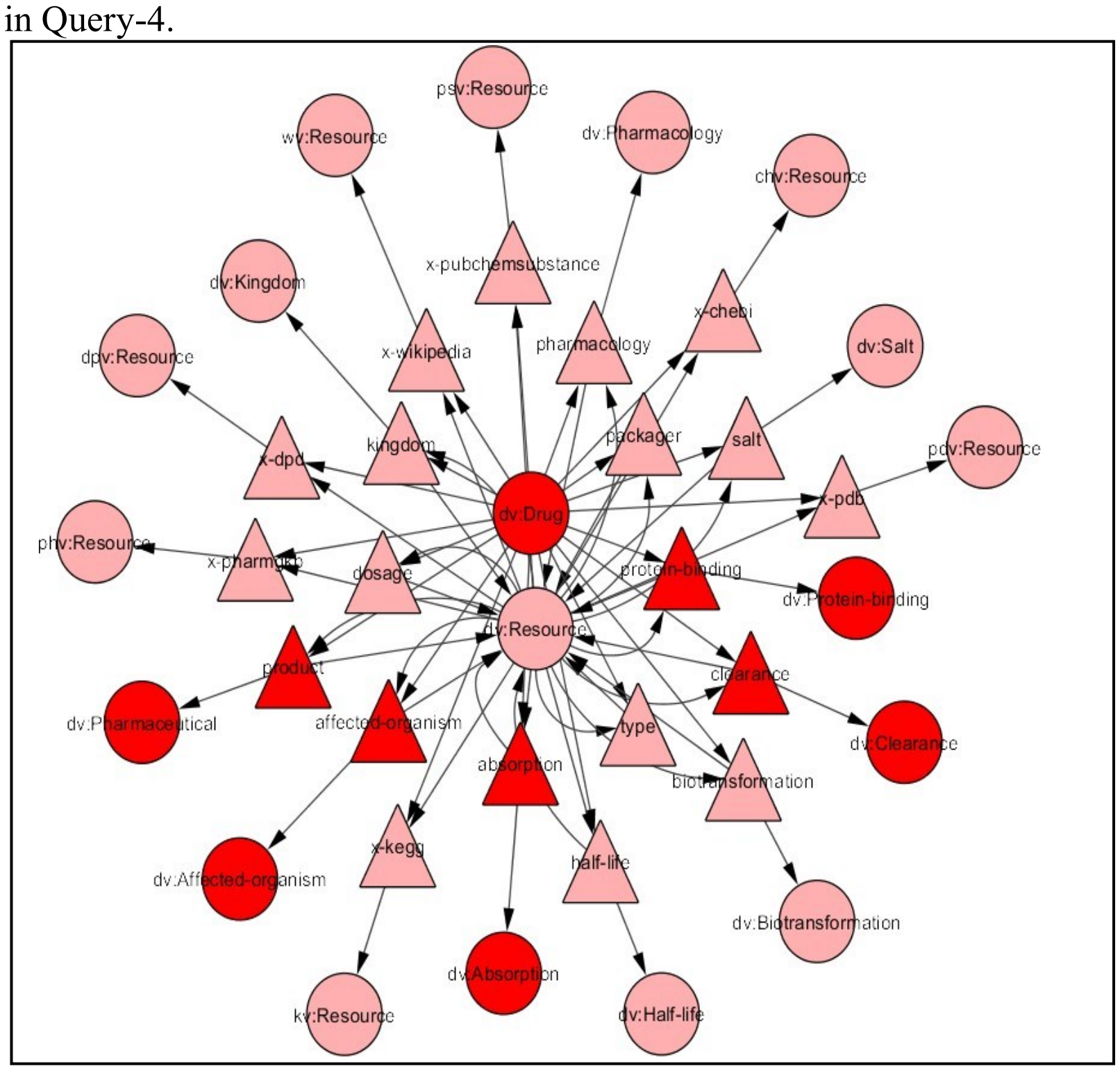

**Figure 17** Topic 3_2 (Topic 2 at Level 3) Graph in DrugBank

**Query-4:** The query graph was automatically generated from Topic 3_2 to depict the topic and query information. This query allows users to find *drugs and their absorption, affected-organism, clearance pharmacokinetic measurement, pharmaceutical information, and protein binding information*. The SPARQL format of Query-4 was automatically generated by the Query Generation algorithm (described in Section II-E) considering the top predicates and their concepts (described in Section IV-B) below.

***Q4. For any two drugs which share the same absorption, affected-organism, clearance and pharmaceutical, what are all the possible combinations?***

```
select distinct ?druglabel, ?drug2label, ?absorptionlabel, ?aolabel, ?clearancelabel, ?pharmaceuticallabel, ?pblabel where {
?drug <http://www.w3.org/1999/02/22-rdf-syntax-ns#type> <http://bio2rdf.org/drugbank_vocabulary:Drug> .
?drug2 <http://www.w3.org/1999/02/22-rdf-syntax-ns#type> <http://bio2rdf.org/drugbank_vocabulary:Drug> .
?drug <http://bio2rdf.org/drugbank_vocabulary:absorption> ?absorption .
Optional {?drug2 <http://bio2rdf.org/drugbank_vocabulary:absorption> ?absorption} .
?drug <http://bio2rdf.org/drugbank_vocabulary:affected-organism> ?ao .
?drug <http://bio2rdf.org/drugbank_vocabulary:clearance> ?clearance .
Optional {?drug2 <http://bio2rdf.org/drugbank_vocabulary:clearance> ?clearance} .
?drug <http://bio2rdf.org/drugbank_vocabulary:product> ?pharmaceutical .
Optional {?drug2 <http://bio2rdf.org/drugbank_vocabulary:product> ?pharmaceutical } .
?drug <http://bio2rdf.org/drugbank_vocabulary:protein-binding> ?pb .
Optional {?drug2 <http://bio2rdf.org/drugbank_vocabulary:protein-binding> ?pb}
?drug <http://www.w3.org/2000/01/rdf-schema#label> ?druglabel.
?drug2 <http://www.w3.org/2000/01/rdf-schema#label> ?drug2label.
?absorption <http://www.w3.org/2000/01/rdf-schema#label> ?absorptionlabel.
?ao <http://www.w3.org/2000/01/rdf-schema#label> ?aolabel.
?clearance <http://www.w3.org/2000/01/rdf-schema#label> ?clearancelabel.
?pharmaceutical <http://www.w3.org/2000/01/rdf-schema#label> ?pharmaceuticallabel.
?pb <http://www.w3.org/2000/01/rdf-schema#label> ?pblabel.
}
```

The Query-4 results on any relevant drugs and their *pharmaceutical* information include {Gemcitabine, Lepirudin, Gemzar 1 gm Solution Vial}, {Tiotropium, Lepirudin, Spiriva 18 mcg Capsule}. As shown in Figure 18, partial query results on Query-4 include the information on some drugs (e.g., Gemcitabine[drugbank:DB00441], Tiotropium[drugbank:DB01409], Carvedilol[drugbank:DB1136]) having the common information of their absorption, affected-organism, clearance pharmacokinetic measurement, pharmaceutical information, and protein binding information.

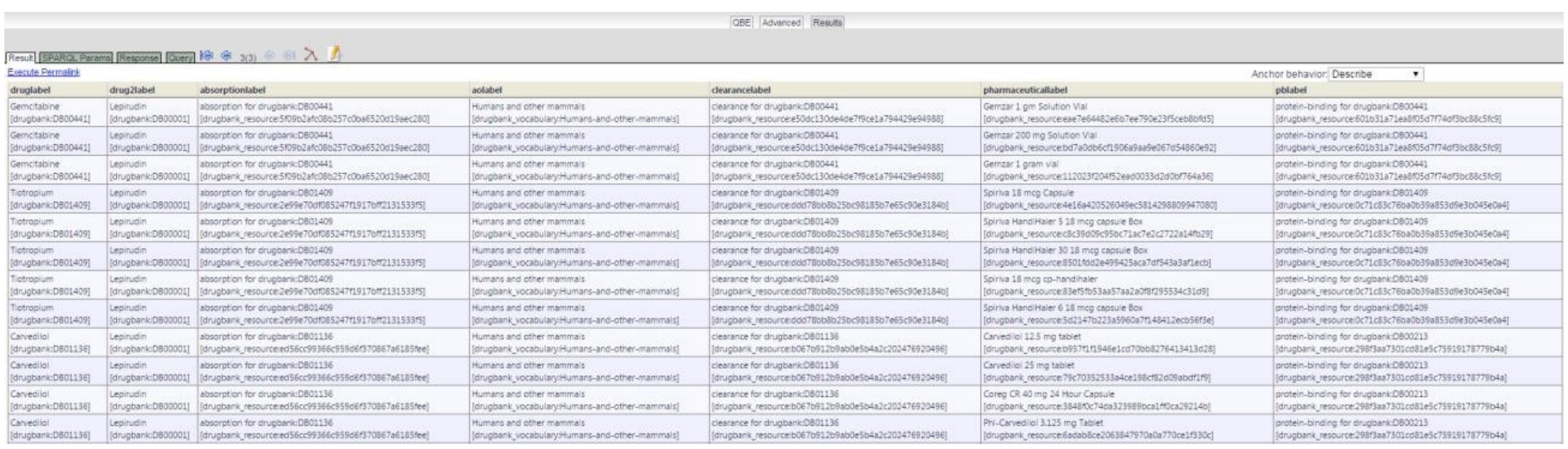

| druglabel | drug2label | absorptionlabel | aolabel | clearancelabel | pharmaceuticallabel | pblabel |
|---|---|---|---|---|---|---|
| Gemcitabine [drugbank:DB00441] | Lepirudin [drugbank:DB00001] | absorption for drugbank:DB00441 [drugbank_resource:5f09b2afc08b257c0ba6520d19aec280] | Humans and other mammals [drugbank_vocabulary:Humans-and-other-mammals] | clearance for drugbank:DB00441 [drugbank_resource:e50dc130de4de7f9ce1a794429e94988] | Gemzar 1 gm Solution Vial [drugbank_resource:eae7e64482e6b7ee790e23f5ceb8bfd5] | protein-binding for drugbank:DB00441 [drugbank_resource:601b31a71ea8f05d7f74df3bc88c5fc9] |
| Gemcitabine [drugbank:DB00441] | Lepirudin [drugbank:DB00001] | absorption for drugbank:DB00441 [drugbank_resource:5f09b2afc08b257c0ba6520d19aec280] | Humans and other mammals [drugbank_vocabulary:Humans-and-other-mammals] | clearance for drugbank:DB00441 [drugbank_resource:e50dc130de4de7f9ce1a794429e94988] | Gemzar 200 mg Solution Vial [drugbank_resource:bd7a0db6cf1906a9aa9e067d54860e92] | protein-binding for drugbank:DB00441 [drugbank_resource:601b31a71ea8f05d7f74df3bc88c5fc9] |
| Gemcitabine [drugbank:DB00441] | Lepirudin [drugbank:DB00001] | absorption for drugbank:DB00441 [drugbank_resource:5f09b2afc08b257c0ba6520d19aec280] | Humans and other mammals [drugbank_vocabulary:Humans-and-other-mammals] | clearance for drugbank:DB00441 [drugbank_resource:e50dc130de4de7f9ce1a794429e94988] | Gemzar 1 gram vial [drugbank_resource:112023f204f52ead0033d2d0bf764a36] | protein-binding for drugbank:DB00441 [drugbank_resource:601b31a71ea8f05d7f74df3bc88c5fc9] |
| Tiotropium [drugbank:DB01409] | Lepirudin [drugbank:DB00001] | absorption for drugbank:DB01409 [drugbank_resource:2e99e70df085247f1917bff2131533f5] | Humans and other mammals [drugbank_vocabulary:Humans-and-other-mammals] | clearance for drugbank:DB01409 [drugbank_resource:ddd78bb8b25bc98185b7e65c90e3184b] | Spiriva 18 mcg Capsule [drugbank_resource:4e16a420526049ec5814298809947080] | protein-binding for drugbank:DB01409 [drugbank_resource:0c71c83c76ba0b39a853d9e3b045e0a4] |
| Tiotropium [drugbank:DB01409] | Lepirudin [drugbank:DB00001] | absorption for drugbank:DB01409 [drugbank_resource:2e99e70df085247f1917bff2131533f5] | Humans and other mammals [drugbank_vocabulary:Humans-and-other-mammals] | clearance for drugbank:DB01409 [drugbank_resource:ddd78bb8b25bc98185b7e65c90e3184b] | Spiriva HandiHaler 5 18 mcg capsule Box [drugbank_resource:c8c39d09c95bc71ac7e2c2722a14fb29] | protein-binding for drugbank:DB01409 [drugbank_resource:0c71c83c76ba0b39a853d9e3b045e0a4] |
| Tiotropium [drugbank:DB01409] | Lepirudin [drugbank:DB00001] | absorption for drugbank:DB01409 [drugbank_resource:2e99e70df085247f1917bff2131533f5] | Humans and other mammals [drugbank_vocabulary:Humans-and-other-mammals] | clearance for drugbank:DB01409 [drugbank_resource:ddd78bb8b25bc98185b7e65c90e3184b] | Spiriva HandiHaler 30 18 mcg capsule Box [drugbank_resource:8501fdd2e499425aca7df543a3af1ecb] | protein-binding for drugbank:DB01409 [drugbank_resource:0c71c83c76ba0b39a853d9e3b045e0a4] |
| Tiotropium [drugbank:DB01409] | Lepirudin [drugbank:DB00001] | absorption for drugbank:DB01409 [drugbank_resource:2e99e70df085247f1917bff2131533f5] | Humans and other mammals [drugbank_vocabulary:Humans-and-other-mammals] | clearance for drugbank:DB01409 [drugbank_resource:ddd78bb8b25bc98185b7e65c90e3184b] | Spiriva 18 mcg cp-handihaler [drugbank_resource:83ef5fb53aa57aa2a0f8f295534c31d9] | protein-binding for drugbank:DB01409 [drugbank_resource:0c71c83c76ba0b39a853d9e3b045e0a4] |
| Tiotropium [drugbank:DB01409] | Lepirudin [drugbank:DB00001] | absorption for drugbank:DB01409 [drugbank_resource:2e99e70df085247f1917bff2131533f5] | Humans and other mammals [drugbank_vocabulary:Humans-and-other-mammals] | clearance for drugbank:DB01409 [drugbank_resource:ddd78bb8b25bc98185b7e65c90e3184b] | Spiriva HandiHaler 6 18 mcg capsule Box [drugbank_resource:5d2147b223a5960a7f148412ecb56f3e] | protein-binding for drugbank:DB01409 [drugbank_resource:0c71c83c76ba0b39a853d9e3b045e0a4] |
| Carvedilol [drugbank:DB01136] | Lepirudin [drugbank:DB00001] | absorption for drugbank:DB01136 [drugbank_resource:ed56cc99366c959d6f370867a6185fee] | Humans and other mammals [drugbank_vocabulary:Humans-and-other-mammals] | clearance for drugbank:DB01136 [drugbank_resource:b067b912b9ab0e5b4a2c202476920496] | Carvedilol 12.5 mg tablet [drugbank_resource:b957f1f1946e1cd70bb8276413413d28] | protein-binding for drugbank:DB00213 [drugbank_resource:298f3aa7301cd81e5c7591917877 9b4a] |
| Carvedilol [drugbank:DB01136] | Lepirudin [drugbank:DB00001] | absorption for drugbank:DB01136 [drugbank_resource:ed56cc99366c959d6f370867a6185fee] | Humans and other mammals [drugbank_vocabulary:Humans-and-other-mammals] | clearance for drugbank:DB01136 [drugbank_resource:b067b912b9ab0e5b4a2c202476920496] | Carvedilol 25 mg tablet [drugbank_resource:79c70352533a4ce198cf82d09abdf1f9] | protein-binding for drugbank:DB00213 [drugbank_resource:298f3aa7301cd81e5c75919178779b4a] |
| Carvedilol [drugbank:DB01136] | Lepirudin [drugbank:DB00001] | absorption for drugbank:DB01136 [drugbank_resource:ed56cc99366c959d6f370867a6185fee] | Humans and other mammals [drugbank_vocabulary:Humans-and-other-mammals] | clearance for drugbank:DB01136 [drugbank_resource:b067b912b9ab0e5b4a2c202476920496] | Coreg CR 40 mg 24 Hour Capsule [drugbank_resource:3848f0c74da323989bca1ff0ca29214b] | protein-binding for drugbank:DB00213 [drugbank_resource:298f3aa7301cd81e5c75919178779b4a] |
| Carvedilol [drugbank:DB01136] | Lepirudin [drugbank:DB00001] | absorption for drugbank:DB01136 [drugbank_resource:ed56cc99366c959d6f370867a6185fee] | Humans and other mammals [drugbank_vocabulary:Humans-and-other-mammals] | clearance for drugbank:DB01136 [drugbank_resource:b067b912b9ab0e5b4a2c202476920496] | Phl-Carvedilol 3.125 mg Tablet [drugbank_resource:6adab8ce2063847970a0a770ce1f330c] | protein-binding for drugbank:DB00213 [drugbank_resource:298f3aa7301cd81e5c75919178779b4a] |

**Figure 18** Partial Results of Query-4 in DrugBank

## V. DISCUSSION

### A. *Knowledge Disocvery from General and Medical Ontologies*

In general settings, many efforts have been made to perform knowledge discovery with semantic web techniques. For instance, the SMARTSPACE proposed a distributed platform for semantic knowledge discovery from services using multi-agent approach[35]. The PEMAR introduced a smart phone middleware for activity recognition discovery based on semantic models[36]. A mobile-cloud computing framework was established to discover infrastructure condition based on a back-end semantic knowledge discovery engine[37]. In our previous work, we have built a situation aware mobile application framework[38, 39] to discovery users' activities in a dynamic way based on the semantic web rule language (SWRL)[40].

In biomedical settings, the advance in Linked Data technologies, such as the standard graph data model (RDF) and distributed SPARQL capability, allows us to easily access distributed data. Most of the work has mainly focused on building or using ontologies for data normalization, bridging and reasoning. Widely used medical ontologies are Bio2RDF[32], TMO (Translational Medicine Ontology)[41], Chem2Bio2RDF[42], SIO (Semanticscience Integrated Ontology)[43], ATC (Anatomical Therapeutic Chemical) and DrugBank integration[44], Linked Life Data[45].

Drug discovery research heavily relies on multiple information sources to validate potential drug candidates as shown in the Open PHACTS project[46]. In complicated domains, it takes time to develop and maintain ontologies[47-50]. There have been various studies on using semantic techniques to improve data integration and share information. DrugBank is one of the key resources which provide bioinformatics and cheminformatics studies with complete information on drug and drug targets. However, these efforts merely support physical integration of multiple biomedical ontologies without considering semantic integration of data. In particular, human intervention is strongly required so that these are not suitable for comprehensive and accurate knowledge discovery especially from a large amount of data. Furthermore, semantic interoperability is difficult to achieve in these systems as the conceptual models underlying datasets are not fully exploited.

Pattern based knowledge analysis has been conducted in many aspects of biomedical research. Warrender and Lord proposed an axiom based generalized and localized pattern driven approach in biomedical ontology engineering[51]. Want *et al*., designed a biomedical pattern discovery algorithm based on a supervised learning approach[52]. Rafiq *et al*., developed an algorithm to discover temporal patterns in genomic databases[53]. However, our approach is different in that, first of all, we focused on a more general approach for graph structural pattern analysis and discovery. In addition, we have combined an unsupervised learning algorithm with a pattern discovery technique to provide a more dynamic way of knowledge discovery from large amount of ontologies.

Our work is motivated by previous work on highlighting the importance of ontological relations[54-56]. Tartir *et al*., pointed out that there are numerous meaningful relations other than class-subclass relations that would be useful for understanding the ontologies[57]. Sabou *et al*., considered ontological relations to be the primary criterion for the summary extraction of ontologies, in which a relatively small number of concepts typically have a high degree of connectivity through hops[58]. In our study, we hypothesized that a similarity measurement based on predicate neighborhood patterns would be more effective in finding relevant information than a concept-based measurement. Our approach defined a new concept of predicate based patterns and neighboring closeness for an automatic knowledge discovery and query generation system.

In this paper, we applied k-means on predicate similarity matrix to group *n* predicates into *k* different clusters in order to form various topics. This approach is unique in terms of three aspects: i) PNP-based similarity measure, ii) a hierarchical approach, iii) level-by-level optimality using a silhouette heuristic function and density function. We utilize the intrinsic property of predicates' neighborhoods, the strong dependence of predicates and their resources with high similarity inside a topic cluster is reflected in their co-occurrence in both their own neighborhoods and the neighborhoods of predicates close-by. It was found that the outcomes from the HPKM algorithm preserve the local neighborhoods of predicates inside topics and shows a high similarity inside a topic cluster (as seen the topic hierarchy in Section IV).

### B. *Query Generation and Query Processing*

There are several works on query generation and processing for ontologies, such as[59-61]. Queries can often be difficult to formulate across these datasets[62]. In particular, the work Lorey et al., proposed[59] has a similar approach to our work in terms of detecting recurring query patterns based on the distance among RDF graph patterns and identifying query templates from the analysis of the RDF graph structure. However, this work focuses more on concepts of the instance level of RDF graphs for the pattern identification and template extraction. Unlike this work, we focused on a new paradigm, such as predicate based similarity patterns, at the schema level for topic discovery and query suggestion.

Biomedical data contributors have provided public SPARQL endpoints to query the datasets. However, studies done by Quilitz et al.,[63] and work developed by Alexander et al.,[64] merely provided the statistical information on the datasets instead of conceptual analysis for knowledge discovery from biomedical datasets. There is little effort for the schema level analysis of the concepts and their relationships in these datasets with respect to systematic and semantic querying. Seaborne and Prudhommeaux pointed out the difficulty with the SPARQL syntax and expression[21], because the precise details of the structure of the graph should be specified for queries in the triple pattern through the various heterogeneous schemas. In reality, users may not be familiar with the details of datasets, and it is hard to express the precise relationships between concepts in the SPARQL syntax and expressions. Thus, this can be a bottleneck for users to query through the endpoints of medical ontologies.

Callahan et al. provided a SPARQLed web application for SPARQL query generation by suggesting context sensitive IRI[60]. However, they could not provide strong associated queries as we do. Unlike this work, we can provide not only valid but also meaningful query suggestions in a dynamic manner according to users' interesting topics. Godoy et al. presented a collaborative environment to allow user to register queries manually through wiki pages and share and execute the queries for linked data[61]. A series of desired queries might be generated using large ontologies like the NCI thesaurus by extracting relevant information[8]. The GLEEN project aims to develop a useful service for simplified, materialized views of complex ontologies[65]. However, these works lack the comprehensive semantic analysis of large sources and the usage of the knowledge for query processing. Unlike these works, our approach is to automate query generation through predicate neighborhood pattern-based topic discovery without any human intervention.

### C. *Future Work on Cross-domain Knowledge Discovery and Query Generation*

In the future, we will apply the proposed approach to heterogeneous biomedical ontologies among multiple domains (e.g., Drug to Gene, Drug to Disease) for knowledge discovery with appropriate semantic granularity[66] and query generation. Specifically, we have investigated the combination of the Human Phenotype Ontology (HPO)[67] with a collaborative filtering algorithm[68] to accelerate rare disease diagnosis[69, 70]. In the future, we plan to incorporate heterogeneous knowledge bases to facilitate the implementation of rare disease diagnosis into clinical practice. The accuracy of the pattern analysis and dynamic similarity computation in cross domain analysis is highly depending on how well datasets are normalized. Therefore, to achieve this goal, the first essential step is to develop an effective normalization technique for heterogeneous biomedical datasets. Famous work in this area include Bio2RDF[32], SIO[43]. In addition to lexicon and semantic based data normalization, some other techniques such as natural language processing and graph pattern based analysis can be used. Word2vec[71] proposed a way to calculate distance between elements based on bag-of-words. By using this approach, we can find synonym terms from heterogeneous domains and normalize terms with highest similarity. LIMES[72] applied triangle inequality on graph data to find the path between a source and target nodes. For a source and target nodes in different datasets, we can find predicates along the path as RDF built-ins <rdfs:sameAs> and <rdfs:seeAlso> and also normalize the synonym terms. A framework was designed to infer the links between entities across multiple heterogeneous social networks[73]. It would be another inspiration for us to better normalize cross domain datasets.

The second task in our future work is to discover more interesting connection patterns and query from integrated datasets since cross domain connection patterns usually provide more valuable heterogeneous information and relationships. There are many data integration works like BioPortal[5], Bio2RDF[32] and OBO[74] which are able to integrate all potential cross domain connection pattern knowledge into one normalized repository. However, none of them has the ability to discover those potential connection patterns in a dynamic way. However, *MedTQ* has a capacity to handle such dynamic cross domain knowledge discovery by exploring new strategies to connect them together and retrieve information from integrated ontologies and data.

## VI. CONCLUSION

In this paper, we proposed the *MedTQ* framework for topic discovery and query generation through the analysis of ontologies. For the *MedTQ* framework, we have newly designed the Predicate Neighboring Pattern (PNP) model and performed similarity measurements, the Hierarchical Predicate-based K-Means clustering (HPKM) algorithm and dynamic query generation algorithm. The proposed *MedTQ* framework was evaluated using a case study with Bio2RDF ontologies (DrugBank). In this case, we demonstrated that *MedTQ* framework can dynamically discover cohesive topics for a given ontology as well as generate interesting queries for the discovered topics. In addition, we successfully validated the optimal clustering results, thereby providing a solid evidence for automatic topic discovery and query generation. In particular, we have implemented and deployed a tool for topic discovery and interactive query generation as well as the SPARQL endpoint for query processing with multiple medical ontologies and datasets from Bio2RDF.

**Figure Legends**

**Figure 1. Predicate Share Patterns**: In a RDF triplet <S, P, O>, there is a direction from a subject S to an object O through a predicate P. From the basic unit of RDF triple <S, P, O>, a specific context of a predicate P can be discovered from the associated concepts (S and O). Interestingly, the neighbors of predicates P will also provide additional information through the association context. In this figure, a circle represents a concept (S stands for a subject and O stands for an object) and a square represents a predicate (relations). The *Share* pattern describes the resources sharing relationships (P) between interacting concepts such as shared subjects (S) or shared objects (O) through the given relationship. This pattern describes that the same subject and object are shared by two predicates $P_1$ and $P_2$ (the leftmost one), the same subject shared (the rightmost one), and the object shared (the middle one).

**Figure 2. Predicate Connection Patterns**: In a RDF triplet <S, P, O>, there is a direction from a subject S to an object O through a predicate P. From the basic unit of RDF triple <S, P, O>, a specific context of a predicate P can be discovered from the associated concepts (S and O). Interestingly, the neighbors of predicates P will also provide additional information through the association context. In this figure, a circle represents a concept (S stands for a subject and O stands for an object) and a square represents a predicate (relations). The *Connection* pattern describes the resources are connected through multiple relationships (P) between interacting concepts such as shared subjects (S) or shared objects (O). The pattern in this figure describes three types of connection patterns: Level 1: given two RDF triples <$S_1$, $P_1$, $O_1$> and <$S_2$, $P_2$ , $O_2$>, $O_1$ is equal to $S_2$ and then there is a predicate connection pattern. Level 2: given three RDF triples <$S_1$, $P_1$, $O_1$>, <$S_2$, $P_2$ , $O_2$>, and <$S_3$, $P_3$, $O_3$>, $O_1$ is equal to $S_2$, $O_2$ is equal to $S_3$, and then there is a predicate connection pattern between $P_1$ and $P_3$. Level 2: given four RDF triples <$S_1$, $P_1$, $O_1$>, <$S_2$, $P_2$ , $O_2$>, <$S_3$, $P_3$, $O_3$>, and <$S_3$, $P_3$, $O_3$>, $O_1$ is equal to $S_2$, $O_2$ is equal to $S_3$, and $O_3$ is equal to $S_4$, ,then there is a predicate connection pattern between $P_1$ and $P_4$.

**Figure 3. Predicate Neighbouring Patterns (PNP) and Similarity Matrix**: This matrix represents the predicate similarity computation for the *Share* patterns and *Connection* patterns. A *Share* pattern is identified between predicates $P_1$ and $P_2$ and the *Connection* patterns are identified between $P_1$ and $P_3$ , $P_1$ and $P_4$, $P_1$ and $P_5$. Based on the PNP patterns, SM$\left[P_i, P_j\right]$ is computed based on the formulas given in Definitions 1 – 4.

**Figure 4. Sublette Width and Number of Topics in Topic Hierarchy**: This figure shows a topic hierarchy that was constructed by the HPKM algorithm. This topic hierarchy is composed of topics in a tree structure. Each node in each topic is representing concepts (circles) and predicates (triangles) of the topic. The average $sw(p_i)$ is computed based on the predicates of each topics (The *sw* value for each topic is shown together with the size of predicates in the parentheses) level-by-level. The branching factor K is defined by the branching optimization approach.

**Figure 5. Automatic Query Generation:** This figure illustrates how to generate a SPARQL query from topics generated. The query generation is based on two step process: for a given topic, i) transforming from a topic graph to a query graph and ii) transforming from the query graph to a SPARQL query. In this example, a topic was discovered from the integration of three ontologies, DrugBank, Sider, PharmGKB ontologies. The query generated from a topic graph includes triples with two predicates, *drug* and *x-pubchem-substance,* and their subject variables (*?E* and *?D*) and object variables *(?D* and *?R*). The type of variable ?E is known as Drug Effect, ?D as Drug, and ?R as PharmGKB Resource. These can be converted to a triplet form such as <?D typeof Drug>. The SPARQL query shown in the top-left textbox is automatically generated for the given topic graph (right-side).

**Figure 6. SPARQL Endpoint:** A web-based query endpoint was developed and deployed. In this example, a SPARQL query was generated from Topic 3_4 of the DrugBank ontology (Topic 4 at level 3) based on the top ranked concepts (such as *dv:Resource*; *dv:Drug*, *dv:Carrier*, *dv:Enzyme*, *dv:Target*) and the top ranked predicates (such as *transporter*, *target*, *enzyme*, carrier) with their concepts *dv:Target-Relation*, *dv:Target-Relation*, *dv:Enzyme-Relation*, *dv:Carrier-Relation*. Through this endpoint, a SPARQL query can be designed in an interactive manner and the query results can be retrieved.

**Figure 7. MedTQ Interactive Query Tool:** This figure shows how to use the *MedTQ* tool for browsing the generated topics and performing interactive design and processing of queries. Step 1 shows the list of topics for a given ontology (DrugBank). Step 2 shows the list of NLP questions for a selected topic (Topic 7). Step 3 shows the automatically generated SPARQL query and the query results. Step 4 shows the topic and query graphs for the selected query.

**Figure 8. DrugBank Topic Hierarchy**: The topic hierarchy was generated for a single domain ontology, *DrugBank*. As seen in this figure, *DrugBank* has the number of topics <2:7:8> with 2 topics at the first level, 7 topics at the second level, and 8 topics at the third level (T3_1, T3_2, T3_3, T3_4, T3_5, T3_6, T3_7, T3_8). K-means clustering was performed in a top-down manner until the average of topics' Silhouette Width is higher than a certain threshold (> 0.5). The number on each edge in the topic hierarchy represents the percentage of predicates that the upper level topic graph contributes to the lower level graph. The eight topics in the bottom level (i.e., level 3) are shown with their high ranked two unique predicates.

**Figure 9. Optimal K Branching Factors using Multiple Clustering Techniques.** Compared to existing clustering algorithms (such as *Clara*, *PAM*, and *Hierarchical Clustering)*, *K-Means Clustering* showed the best optimal cluster number showing with a stable silhouette width for varying the number of clusters. The proposed HPKM algorithm, which is an extension of K-Means clustering, determines the most significant number of branching factors for clustering at each level (shown as a red circle) such as (a) K = 2 with SW = 0.77 at level 1; (b) K = 5 with SW = 0.64 at level 2; (c) K=2 with SW = 0.74 at level 2; (e) K = 2 with SW = 0.72 at level 3.

**Figure 10. Topic Rankings for DrugBank**. This shows relevance scales of five different rankings and an overall ranking. The overall rank was computed in terms of the five criteria: i) *Top 20 Concepts*, ii) *Top 20 Predicates*, iii) *Similarity*, iv) *Silhouette Width*, v) *Density*. The numbers shown in this figure are rankings. Topic 4 at level 3 (T3_4) and Topic 7 at level 3 (T3_7) are ranked 1st and 2nd (of 8 topics), respectively.

**Figure 11. Topic 3_4 (Topic 4 at Level 3) Graph in DrugBank**. This topic graph includes 6 predicates and 12 concepts with 72 in-degree and out-degree. Specifically, two groups of predicates are shown; one is with four predicates such as *transporter*, *target*, *enzyme*, *carrier* with concepts *dv:Target-Relation*, *dv:Target-Relation*, *dv:Enzyme-Relation*, *dv:Carrier-Relation*. Another group of predicates such as *x-genbank* and *x-uniprot* is a connector predicate group that connects between internal concepts (e.g., *dv:Drug*, *dv:Enzyme*) and external concepts (e.g., *gv:Resource*, *unv:Resource*). In this graph, concepts are represented as a circle, predicates as a triangle, and links as an arrow. Specifically, the dark red items are the predicates and concepts mentioned in Query-1. This topic describes the relations between drugs with their target, enzyme, enzyme relation, and target relation. The prefixes in this graph describe the domain of the concepts as follows: dv:http://bio2rdf.org/drugbankvocabulary and gv: http://bio2rdf.org/genbankvocabulary

**Figure 12. Partial Results of Query-1 in DrugBank.** This figure shows the partial results from Query-1 about drug target and transporter enzyme of some drugs (e.g., Gemcitabine [drugbank:DB00441], Fluorouracil [drugbank:DB00544], Ribavirin [drugbank:DB00811]). All the possible drugs, enzyme, target, enzyme relations, target relations for any drugs that share the same target and transporter enzyme.

**Figure 13. Topic 3_7 (Topic 7 at Level 3) Graph in DrugBank.** Topic 3_7 is highly ranked. In this topic graph, three predicates, (i.e., *drug*, *action*, *reference*) are all nicely connected with 7 concepts (dv:Transporter-Relation, dv:Target-Relation, dv:Drug, dv:Carrier-Relation, dv:Enzyme-Relation, pv:Resource, dv:Resource). Concepts are represented as a circle, predicates as a triangle, and links as an arrow. In addition, the dark red triangle items (*drug* and *reference*) are the predicates and the dark red circle items (*dv:Transporter-Relation*, *dv:Target-Relation*, *dv:Drug*, *dv:Carrier-Relation*) are the concepts mentioned in Query-2. The prefixes in this graph describe the domain of the concepts as follows: dv: http://bio2rdf.org/drugbankvocabulary and pv: http://bio2rdf.org/pubmedvocabulary

**Figure 14. Results of Query-2 in DrugBank.** This figure shows the partial results from Query-2 about some drugs such as *Phenytoin* (DrugBank:DB00252), *Lepirudin* (DrugBank:DB00001) and *Deferasirox* (DrugBank:DB01609). All the possible combinations and their PubMed references for *Target-relation*, *Carrier-relation* and *Transporter-relation* of drugs sharing the common information of *Target-relation*, *Carrier-relation* and *Transporter-relation* are retrieved.

**Figure 15. Graph of Query-3 in DrugBank**. This shows T3_6 (Topic 6 at level 3) topic graph in which concepts are represented as a circle, predicates as a triangle, and links as an arrow. In particular, there are two subgraphs; one is with two predicates such as *source* and *calculated-properties* with concepts *dv:Boiling-Point* and *dv:Bioavailability*, respectively. Another predicate *experimental-properties* is connected with concepts such as

*dv:Water-Solubility*. The prefix in this graph describes the domain of the concepts and predicates as follows: dv: http://bio2rdf.org/drugbankvocabulary.

**Figure 16. Results of Query-3 in DrugBank**. Query-3 is designed to retrieve information on some drugs' experimental properties and calculated properties of drugs that contain octanol-water partition coefficient. The partial query results on Query-3 include the information on some drugs (e.g., L-Histidine [drugbank:DB00117], L-Phenylalanine [drugbank:DB00120], L-Arginine [drugbank:DB00125]) and all their experimental properties and calculated properties that have *LogP experimental properties (octanol-water partition coefficient)*.

**Figure 17. Topic 3_2 (T3_2) Graph in DrugBank.** In this graph, concepts are represented as a circle, predicates as a triangle, and links as an arrow. In this graph, two dominant concepts, *dv:Resource* and dv:*Drug*, are fully connected to the remaining concepts via predicates such as *absorption*, *protein-binding*. In this graph, concepts are represented as a circle, predicates as a triangle, and links as an arrow. The dark red items like *abortion* and *product* are the predicates and *dv:Drug* and *dv:Pharmaceutical* are the concepts mentioned in Query-4. The dark red triangle items like *abortion* and *product* are the predicates and the dark red circle items like *dv:Drug* and *dv:Pharmaceutical* are the concepts mentioned in Query-4. The prefixes in this graph describe the domain of the concepts as follows: dv: http://bio2rdf.org/drugbankvocabulary, chv: http://bio2rdf.org/chebivocabulary, dpv: http://bio2rdf.org/dpdvocabulary, kv: http://bio2rdf.org/keggvocabulary, pdv: http://bio2rdf.org/pdbvocabulary, phv: http://bio2rdf.org/pharmgkbvocabulary, psv: http://bio2rdf.org/pubchem.substancevocabulary, and wv:http://bio2rdf.org/wikipediavocabulary

**Figure 18. Partial Results of Query-4 in DrugBank.** Query-4 is designed to retrieve information on some drugs' *absorption, affected-organism, clearance pharmacokinetic measurement, pharmaceutical information, and protein binding information*. The partial query results on Query-4 include the information on some drugs (e.g., Gemcitabine[drugbank:DB00441], Tiotropium[drugbank:DB01409], Carvedilol[drugbank:DB1136]) having the common information of their absorption, affected-organism, clearance pharmacokinetic measurement, pharmaceutical information, and protein binding information.